\documentclass[fleqn,usenatbib]{mnras}

\usepackage{newtxtext,newtxmath}

\usepackage[T1]{fontenc}

\DeclareRobustCommand{\VAN}[3]{#2}
\let\VANthebibliography\thebibliography
\def\thebibliography{\DeclareRobustCommand{\VAN}[3]{##3}\VANthebibliography}

\usepackage{graphicx}	
\usepackage{amsmath}	
\usepackage{xcolor}
\usepackage{enumitem}

\newcommand{\ff}{{\sc firefly}}
\newcommand{\ffvac}{\texttt{MaNGA {\sc firefly} VAC}}

\title[Stellar Populations in Barred MaNGA Galaxies]{Azimuthal Variations of Stellar Populations in Barred Galaxies}

\author[J. Neumann et al.]{
Justus Neumann,$^{1,2}$\thanks{E-mail: jusneuma.astro@gmail.com}
Daniel Thomas$^{2,3}$
Claudia Maraston,$^{2}$
Damian R. Gleis,$^{1}$
Chuanming Mao,$^{1}$\newauthor
Eva Schinnerer,$^{1}$
Sophia K. Stuber,$^{1}$
\\
$^{1}$Max-Planck-Institut f{\"{u}}r Astronomie, K{\"{o}}nigstuhl 17, D-69117 Heidelberg, Germany\\
$^{2}$Institute of Cosmology and Gravitation, University of Portsmouth, Burnaby Road, Portsmouth PO1 3FX, UK\\
$^{3}$School of Mathematics and Physics, University of Portsmouth, Lion Gate Building, Portsmouth PO1 3HF, UK
}

\date{Accepted 2024 September 25. Received 2024 September 13; in original form 2024 June 27}

\pubyear{2024}

\begin{document}
\label{firstpage}
\pagerange{\pageref{firstpage}--\pageref{lastpage}}
\maketitle

\begin{abstract}

    Bars are expected to impact the distribution of stellar populations both during bar formation, as they rearrange stars into new orbits, and afterwards, due to the redistribution of star-formation-fuelling gas and transfer of angular momentum. We study the impact of stellar bars on the azimuthal variation of stellar population age, metallicity and mass surface density in $\sim1\,000$ nearby barred galaxies from the SDSS-IV/MaNGA survey. Bars have higher stellar mass density ($0.113^{+0.065}_{-0.067}$\,dex) and are more metal-rich ($0.028^{+0.033}_{-0.040}$\,dex) than the discs at the same radii. Stellar ages show a variety of bar to inter-bar contrasts with no consistent trend. The difference in metallicity increases with total stellar mass of the galaxy and distance below the star-forming main sequence. We discuss a combination of potentially responsible processes including kinematic separation, more extended star formation histories and more efficient recycling in bars and at bar-spiral arm connections. Additionally, we observe an offset ($10\degr$-$40\degr$) of the peak metallicity to the bar major axis in star-forming bars in low-mass galaxies, and more metal-rich regions outside the ends of the bar in long bars and quenched galaxies. Furthermore, there is a subtle trend of lower metallicities on the leading side of spiral arms compared to the trailing side. Finally, we report a spiral arm surface density feature, which could point towards a dominant bar-spiral connection and pitch angle of $\alpha \sim 25\degr$. We interpret these features in the context of bar formation and the impact of large-scale gas flows associated with their presence.

\end{abstract}




\section{Introduction}

Galactic bars are elongated stellar structures in the central part of 60-70\% nearby disc galaxies \citep{Eskridge2000,Menendez-Delmestre2007,Sheth2012a,Simmons2014a,Erwin2018}. They are often prominent morphological features and play a major role in secular galaxy evolution. Frequently, they are described as the \emph{engine} or the \emph{main driver} of secular evolution \citep[e.g.][]{Kormendy2004}. Torques introduced by the non-axisymmetric nature of a barred gravitational potential redistribute angular momentum and matter in galaxy discs \citep{Lynden-Bell1972,Combes1985,Athanassoula2003,Athanassoula2005}. Thereby, bars drive the formation of secularly-built structures, such as inner and outer rings \citep{Buta1986,Buta1996}, nuclear rings or discs \citep{Knapen1995,Debattista2007,Fragkoudi2019a,Bittner2020,Gadotti2020}, and nuclear (or inner) bars \citep{Lorenzo-Caceres2012,Lorenzo-Caceres2013}.

While bars have been studied since decades, for a long time they were thought to be short-lived, rather transient features \citep{Lynden-Bell1972,Friedli1993,Bournaud2005}. Over the last two decades, results both from observations and simulations seem to converge that bars started to form early in the universe. Recent studies based on data from the James Webb Space Telescope (JWST) find bars in galaxy as early as 2$\,$Gyr after the Big Bang \citep{Guo2023,Costantin2023,LeConte2024,Guo2024} and using submillimeter observations further evidence was reported for potential bar formation when the Universe was even only $\sim 0.8\,$Gyr old (\citealp{Amvrosiadis2024}, see also \citealp{Smail2023}, \citealp{Tsukui2024}, and using simulations from TNG50, \citealp{RosasGuevara2022}).
In addition, bars have been shown to be typically long-lived \citep{Gadotti2015,RosasGuevara2020,Fragkoudi2021, deSaFreitas2023a,deSaFreitas2023b} and hard to destroy \citep{Shen2004,Athanassoula2005,Berentzen2007}. Hence, it is now very clear that bars are fundamental for the understanding of galaxy evolution.

Stars are archaeological relics that allow us to trace back the history of mass assembly and chemical enrichment in galaxies. While observed gas properties provide us with information about the current chemical state of a galaxy, stars encode evidence of a galaxy's past baryonic cycles. Stellar population measurement in external galaxies are indispensable to constrain our understanding and development of models and simulations of galaxy evolution \citep[][and references therein]{Thomas2005,Thomas2010,Parikh2021}.

A galactic bar is formed out of stars on mostly quasi-periodic orbits, dominated by the so-called $x_1$ orbits, which are elongated along to the bar major axis and build the backbone of the bar \citep{Contopoulos1980,Athanassoula1983,Pfenniger1984,Skokos2002,Skokos2002a}. Since most stars in the bar will stay on these orbits rotating with a common pattern speed, they are well separated from the rest of the disc over a long time period (not accounting for mergers, radial migration or newly trapped stars). An analysis of the ages and chemical composition of the stellar population of the bar in comparison with those of the background disc will help us to understand the evolution of the bar itself and the influence of the bar on the host galaxy. For example, we are able to learn about separation of stellar populations during bar formation, as well as the efficiency of star formation, quenching and chemical enrichment in the inner galaxy.

Early observational work on stellar populations of bars have focused on radial gradients along the bar major axis compared to the minor axis or the inner or outer disc \citep[e.g.][]{Perez2007,Perez2009,Perez2011,Sanchez-Blazquez2011,Sanchez-Blazquez2014,Seidel2016,Fraser-McKelvie2019}. While still partially ambiguous, these studies provide evidence for a flattening of stellar metallicity gradients along the bar major axis as compared to a steeper decrease in the disc, in support of the idea that bars are more metal-rich than discs. Flatter gradients are in agreement with theoretical expectations of orbital mixing \citep{Binney1987} and radial migration \citep{Sellwood2002,Minchev2010,DiMatteo2013,Grand2012,Grand2015,Halle2015,Halle2018}. Only very recently some attention has been drawn to explore in a fuller extent the 2D information provided by resolved stellar population analysis in external barred galaxies \citep{Neumann2020}.

In \citet{Neumann2020}, we explore stellar age, [Z/H] and [Mg/Fe] abundance maps, as well as the star formation histories (SFHs) in the centres of nine nearby barred galaxies as part of the Time Inference with MUSE in Extragalactic Rings \citep[TIMER;][]{Gadotti2019} project using integral-field unit (IFU) data from the
Multi-Unit Spectroscopic Explorer \citep[MUSE;][]{Bacon2010}
on the Very Large Telescope (VLT). The results of this analysis are that bars are on average more metal-rich and less [Mg/Fe]-enhanced than the surrounding discs, which we interpret as the result of bar-induced star formation quenching in the inner disc. Furthermore, we detect a characteristic V-shaped signature in the SFH across the width of the bars, indicative of younger stars being on more elongated orbits closer to the bar major axis. We compare this to the Auriga cosmological zoom-in simulations \citep{Grand2017,Fragkoudi2020} and find that this V-shape arises as a consequence of kinematic separation of stellar populations during bar formation \citep[see also][]{Athanassoula2017,Debattista2017,Fragkoudi2017}.

Our MUSE study of bars in TIMER provided valuable insights into the evolution of bars and their host galaxies, but was limited to only nine massive ($M_\star > 10^{10.4}\,M_\odot$) galaxies. In the present work, we build on these results by expanding our research to the largest IFU survey of nearby galaxies to date, the Mapping Nearby Galaxies at Apache Point Observatory survey \citep[MaNGA;][]{Bundy2015}. We analyse stellar population maps of $\sim 1\,000$ barred galaxies and specifically focus on the azimuthal variations of stellar metallicity, age and mass surface density. These data allows us to probe trends across different masses, star formation rates (SFRs) and bar properties in nearby galaxies with statistical significance.

The paper is organised as follows: In Sect. \ref{sect:data}, we present the MaNGA survey, our stellar population catalogue, the Galaxy Zoo: 3D programme and our sample selection. Next, in Sect. \ref{sect:analysis}, we describe our analysis that transforms single galaxy Voronoi-binned stellar population maps to sample-averaged azimuthal variation polar plots. Subsequently, in Sect. \ref{sect:results}, we present our main results and how they vary across different subsamples. It follows a discussion in Sect. \ref{sect:discussion} and a summary with concluding remarks in Sect. \ref{sect:conclusion}.

\section{Data}
\label{sect:data}

    There are two essential input quantities for this work: spatially resolved stellar population maps and bar measurements. We gain the first from spectral model fitting of optical IFU data, while for the second we use a two-fold approach with a combination of bar identification via crowdsourcing and geometric fitting. In the following, we first present the IFU galaxy survey in Sect. \ref{sect:manga}, the catalogue of stellar population parameters in Sect. \ref{sect:vac}, the citizen science catalogue in Sect. \ref{sect:gz3d} and the sample selection in \ref{sect:sample}.

    \subsection{MaNGA Survey}
    \label{sect:manga}

        The Mapping Nearby Galaxies at Apache Point Observatory survey \citep[MaNGA;][]{Bundy2015} is currently the largest IFU survey of nearby galaxies. It observed $10\,010$ unique galaxies across $4\,000 \deg^2$ at a median redshift of $z\sim 0.037$. The fully reduced dataset alongside with high-level data products from the MaNGA data reduction pipeline \citep[\texttt{DRP;}][]{Law2016} and the MaNGA data analysis pipeline \citep[\texttt{DAP;}][]{Westfall2019} were released to the public in \citet{Abdurrouf2021}.  

        MaNGA is a Sloan Digital Sky Survey-IV project \citep[SDSS-IV;][]{Blanton2017} that uses multiple hexagonal IFU fibre bundles plugged into observation plates that feed into the BOSS spectrographs \citep{Smee2013} mounted at the Sloan Foundation 2.5-meter Telescope \citep{Gunn2006}. A uniform galaxy coverage out to 1.5 $\times$ effective radius ($R_\mathrm{e}$) for the \emph{Primary Sample} and 2.5 $\times\,R_\mathrm{e}$ for the \emph{Secondary Sample} is accomplished by matching galaxy diameters to variable IFU sizes \citep{Drory2015}.

        The spectrographs consist of a red and a blue camera that cover the full wavelength range between $3622\,$\AA$\,$ and $10354\,$\AA$\,$ at a median spectral resolution of $\rm \sigma = 72\,km\,s^{-1}$. The median angular resolution is $2.54\,\arcsec$ full width at half maximum (FWHM), which corresponds to $1.8\,\mathrm{kpc}$ at the median redshift of $z\sim0.037$ \citep{Law2016}. With $\sim 2$-$3\,\mathrm{h}$ integration time, the observations typically have a signal-to-noise ratio (S/N) of $\sim $5-10 per pixel in the $r$-band at 1.5$\,R_\mathrm{e}$ \citep{Yan2016b}.
    
    \subsection{MaNGA Firefly Value-Added-Catalogue}
    \label{sect:vac}

        Sloan surveys have made it a tradition to share and promote so-called \emph{Value-Added-Catalogues} (VACs), which are archives of secondary data products built by different groups from within the collaboration and released to the public. As part of the MaNGA survey, we have built the \ffvac\footnote{\url{https://www.sdss.org/dr17/manga/manga-data/manga-firefly-value-added-catalog}}: a catalogue of global and resolved stellar population parameters for all $10\,010$ galaxies of the final MaNGA data release (DR) in DR17 of SDSS-IV \citep{Abdurrouf2021}. The VAC is fully presented in \citet{Neumann2022} with some further applications and performance testing of earlier release versions in \citet{Goddard2017} and \citet{Comparat2017}. Here, we summarise some of its main features.

        We use \ff\footnote{\url{https://www.icg.port.ac.uk/firefly/}} (\citealp{Wilkinson2017}, with an update presented in \citealt{Neumann2022}) -- a full spectral fitting code of stellar populations -- to fit weighted linear combinations of simple stellar population models (SSPs) to observed MaNGA IFU spectra. The \ffvac{} builds on the \texttt{DAP} and complements its stellar kinematic \citep{Westfall2019} and emission line analysis \citep{Belfiore2019} with stellar population parameters. As part of the {\sc dap} all reduced datacubes are spatially binned using the Voronoi tessellation method \citep{Cappellari2003} to a minimum target S/N$\sim$10. Afterwards, the {\sc dap} employs \texttt{pPXF} \citep{Cappellari2004,Cappellari2017} to fit the stellar kinematics of the binned spectra with a hierarchically clustered selection of stellar templates from the MILES library \citep{Sanchez-Blazquez2006}. Details of the emission line module can be found in \citet{Belfiore2019}. The Voronoi-binned datacubes with emission line fits and stellar kinematic parameters are the main input for the \ffvac.

        Each emission line-subtracted spectrum per Voronoi bin is fitted by \ff{} twice using two different SSP model libraries: the \texttt{M11-MILES} model templates from \citet{Maraston2011} and a new version of the \texttt{MaStar SSP} models described in \citet[][and in prep.]{Maraston2020} based on the MaStar stellar library \citep{Yan2019}. In both cases a \citet{Kroupa2001} IMF is assumed. Further details about the models, theoretical assumptions and the \ff{} code can be found in \citet{Neumann2022}. The \ffvac{} in its two variants \texttt{FF-Mi} and \texttt{FF-Ma} contains twice a total of $>3.7$ million stellar population parameter sets per Voronoi bin across $\sim $10,000 galaxy observations with an additional set of global parameters per galaxy. In this work we make use of the resolved maps of light-weighted stellar age, metallicity and mass surface density in the \texttt{FF-Mi} variant. Most of our main results do not change when we use the \texttt{FF-Ma} variant instead, with the exception of stellar ages as discussed in Sect. \ref{Sect:trendSSP}. Using mass-weighted parameters instead of light-weighted averages does not change our results.

    \subsection{Galaxy Zoo: 3D}
    \label{sect:gz3d}

        Galaxy Zoo: 3D (hereafter GZ:3D, \citealp{Masters2021}) is a citizen science crowdsourcing project that followed in the footsteps of the successful Galaxy Zoo \citep{Lintott2011}. While Galaxy Zoo was aiming at obtaining basic morphological classifications of galaxies, with more detailed classifications in Galaxy Zoo 2D (GZ 2D, \citealp{Willett2013}), GZ:3D asked volunteers to visually identify and draw the outlines of some of these morphological structures, such as spiral arms and bars. At least 15 volunteers per galaxy were necessary to count it as successful. A final ``average'' drawing can then be obtained by applying a vote fraction, which counts -- for every pixel -- how many participants voted that pixel to be within the outline of the structure. GZ:3D obtained classifications for 9188 galaxies in the MaNGA sample. In this work, we use the bar drawings for all barred galaxies in that sample\footnote{The presence of a bar is obtained from a pre-selection based on GZ 2D where at least $20\%$ of people saw a bar.}, which amounts to 1355 galaxies.
    
    \subsection{Sample}
    \label{sect:sample}

        \subsubsection{Main selection criteria and sample characteristics}

        \begin{figure}
    	\centering
    	\includegraphics[width=\columnwidth]{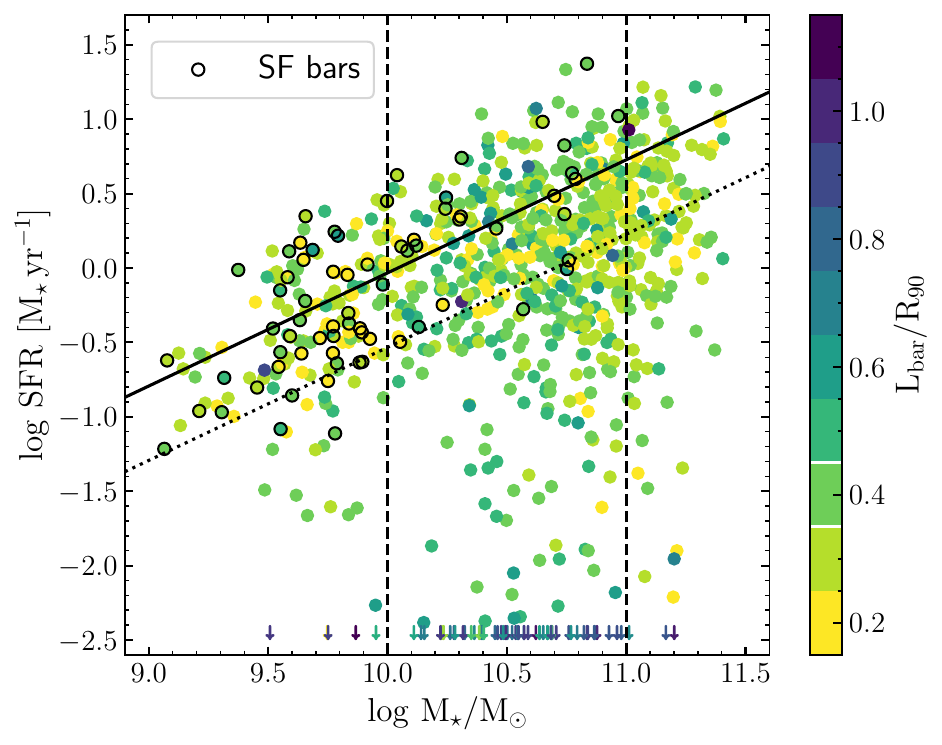}
            \caption{Sample representation in total SFR versus total stellar mass plane. SFRs are based on dust-corrected H$\alpha$ measurements from \texttt{Pipe3D} \citep{Sanchez2022}. Points are coloured by scaled bar length with L$_\mathrm{bar}$ as described in Sect. \ref{sect:lbar} and R$_{90}$ from the enhanced NASA Sloan Atlas \citep{Wake2017}. Star-forming bars are marked using the information provided in \citet{FraserMcKelvie2020a}. The black solid line marks the star-forming main sequence (SFMS) as determined by \citet{Renzini2015} and the dotted line 0.5\,dex below the solid line represents our selection of galaxies on or below the main sequence. The two dashed vertical lines illustrate our subsamples separated in mass and the two white horizontal lines on the colourbar show our sample separation by bar length. Points outside the y-axis range are illustrated as arrows at the bottom.}
            \label{fig:sample}
        \end{figure}
    
        \begin{figure}
    	\centering
    	\includegraphics[width=\columnwidth]{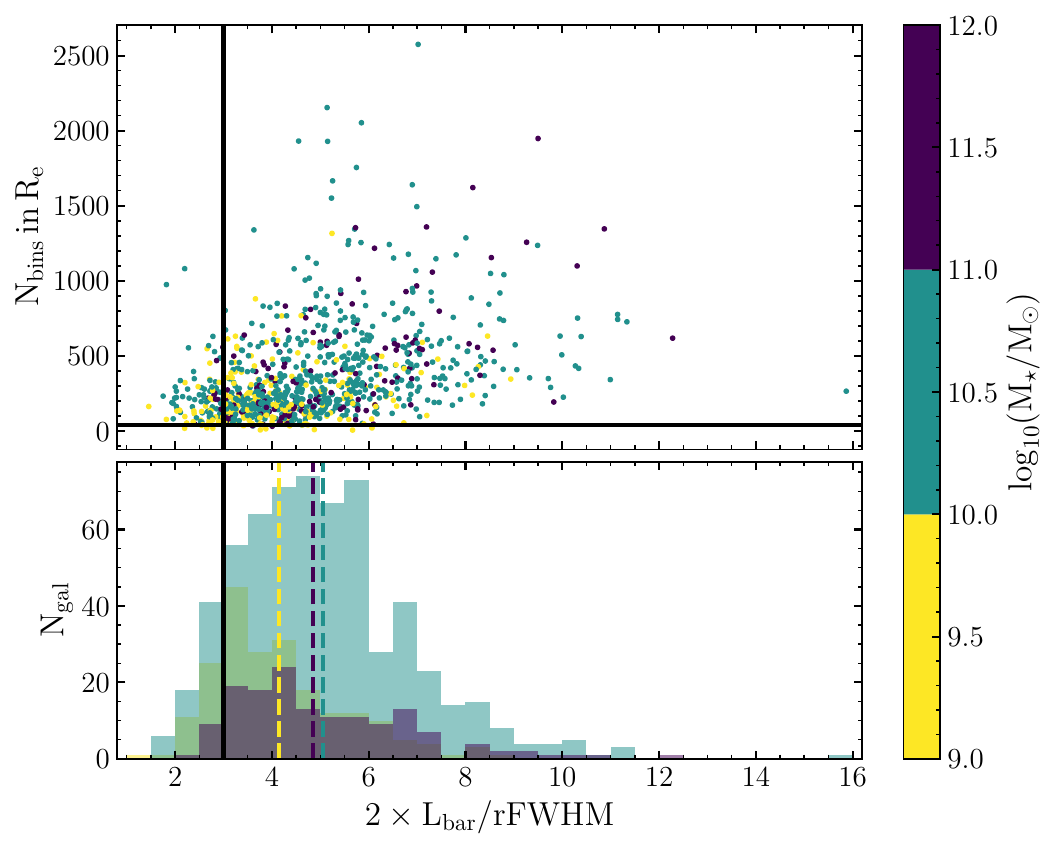}
            \caption{Illustration of spatial resolution and spatial sampling of the sample of 977 galaxies with bar length measurements. The x-axis shows the number of resolution elements, measured as FWHM in the $r$-band, per full bar extent. The y-axis in the top panel shows the number of Voronoi bins within the effective radius $R_{\rm e}$. The bottom panel presents histograms dividing the sample in the same three stellar mass bins as in Fig. \ref{fig:sample}. The solid lines mark the cuts we applied in our selection of the final 846 galaxies into our working sample, as discussed in the text. The vertical dashed lines show the medians after applying the cuts.}
            \label{fig:sample_res}
        \end{figure}
    
            The main requirement for our sample selection is the clear presence of a bar, whereby we are likely to bias our sample to strong bars \citep[i.e. SB types,][]{deVaucouleurs1991,Masters2011,Geron2021}. Morphological classifications including a search for bars for (almost) all MaNGA galaxies have been obtained visually from expert classifiers in \citet{VazquezMata2022}, from GZ 2D crowdsourcing in \citet{Willett2013} and via deep-learning in \citet{DominguezSanchez2022}.
    
    
            We choose to use the classifications from GZ 2D given the availability of bar masks for most of them in GZ:3D. Thus, the parent sample in this work are all barred MaNGA galaxies with available bar masks, namely 1355 galaxies. All but one of these galaxies have available \ff{} maps. One galaxy had low-quality data and was not processed by {\sc dap} and \ff. We carefully inspected all galaxies visually and removed galaxies that were (1) clearly interacting or merging, (2) having foreground stars that impact the stellar population analysis of bar recognition, or (3) where bar drawings were clearly problematic or wrong. A combination of (1) and (3) was the most frequent cause for removal. This selection left us with $1\,038$ galaxies. We further removed all galaxies with high inclination ($i>65\degr$) based on the inclinations given in \citet{Neumann2021} leaving us with 977 galaxies. This is the sample used for the bar length measurement in Sect. \ref{sect:lbar}. Our main results do not change, if we choose a more restrictive inclination limit, as discussed in Appendix \ref{apx:test_incl}. Finally, for further stellar population analysis and in order to secure sufficient spatial resolution and sampling, we further limit our sample to galaxies with at least three resolution elements (FWHM) per full bar extent (twice the bar radius) and 40 Voronoi bins within $R_{\mathrm{e}}$. We are left with our final working sample for the stellar population analysis of 846 galaxies.
            
            Figure \ref{fig:sample} presents our final sample in the SFR vs. stellar mass plane. SFRs are taken from \citet{Sanchez2022} and are based on extinction-corrected H${\alpha}$ fluxes. Stellar masses are from the global properties in the \ffvac{} and are based on photometric measurements in the updated NASA Sloan Atlas \citep[NSA,][]{Wake2017}.  Subsample definitions are explained in the next subsection.

            In Fig. \ref{fig:sample_res}, we illustrate how well bars in our sample are resolved and sampled. Our selection removes $\sim 1\%$ of poorly sampled galaxies and $\sim 12\%$ of poorly resolved bars. While low-mass galaxies host on average smaller and less-well resolved bars, the difference is small (4.1 resolution elements per bar compared to 5.0 and 4.8 for intermediate- and high-mass galaxies) and the distributions are not altered significantly by the cut applied in our selection. Increasing the required minimum resolution does not change the main trends of our results but decreases the azimuthal differences with increasing spatial smearing and shows less significance as the number statistics go down. A study with higher resolution data would potentially show stronger differences compared to those presented here.

        \subsubsection{Subsample definition}
        \label{sect:subsamples}

            This work makes use of the large sample size to investigate trends of the derived parameters along a number of galaxy properties. In Sect. \ref{sect:results}, we report trends with galaxy total stellar mass, distance from the SFMS, SF inside the bar and length of the bar. The chosen demarcation between low-mass ($\log_{10} (M_\star/M_\odot) <10$), intermediate-mass and high-mass galaxies ($\log_{10} (M_\star/M_\odot) \geq 11$) are shown in Fig. \ref{fig:sample} with vertical dashed lines. Star-forming and quiescent galaxies are separated in the same figure by the dotted line 0.5\,dex below the SFMS. For a subsample of galaxies, we obtained from \citet{FraserMcKelvie2020a} a classification of the presence of star formation inside the bar. In that work, the morphology of H$\alpha$ emission in barred galaxies of the MaNGA survey is classified into different categories, including a class of star-forming bars, while all other categories, with star formation only in the centre, at the bar ends or no star formation at all, can be bundled into a class of non-star-forming bars. Since most of the star-forming bars were identified in low-mass galaxies, we limit our non-star-forming bar comparison sample to the same mass range, i.e. $\log M_\star / M_\odot <10$. Finally, we separate our sample into short ($L_\mathrm{bar}/R_{90} < 0.35$), medium-sized ($0.35<L_\mathrm{bar}/R_{90}<0.45$) and long bars ($L_\mathrm{bar}/R_{90}>0.45$), illustrated by the colourbar of Fig. \ref{fig:sample}.

\section{Analysis}
\label{sect:analysis}

    Our goal in this work is to measure azimuthal variations of stellar populations in barred galaxies. In particular, we are interested in how stellar populations change between the bar and inter-bar region at fixed radius. We aim for a statistical analysis using large samples of galaxies rather than individual objects. This will enable us not only to formulate broader statements about the local galaxy population but also to mitigate some of the caveats of a large survey such as MaNGA in terms of spatial resolution and S/N. To achieve this, we stack stellar population maps of galaxies. Note that we do not stack the raw spectra themselves, but the high-level stellar population measurements such as stellar age, metallicity and mass density. One could argue that stacking the spectra directly prior to the stellar population analysis is favourable, as it would increase the S/N of the spectrum itself and allow for a better spectral fit. However, stacking spectra from potentially very different populations across galaxies unavoidably drags in biases (stacking spectra makes sense for uniform populations such as those in elliptical galaxies, much less so in spirals or any star-forming galaxy). Therefore, we opt for using the existing stellar population maps and performing the stacking a posteriori, which also effectively increases the S/N of the final product. It is important to point out that the stellar population measurements were performed using the projected line-of-sight observations. The pixel positions of the derived stellar population parameters are then deprojected before stacking, as described in Sect. \ref{sect:analysis_stacking}. 

    Before we can stack the maps, we need to align them in a sensible way. For the alignment, we require the bars to be on top of each other and, hence, we need to know the position angle and the length of each bar.

    \subsection{Bar length measurement}
    \label{sect:lbar}

    \begin{figure*}
        \centering
        \includegraphics[width=\linewidth]{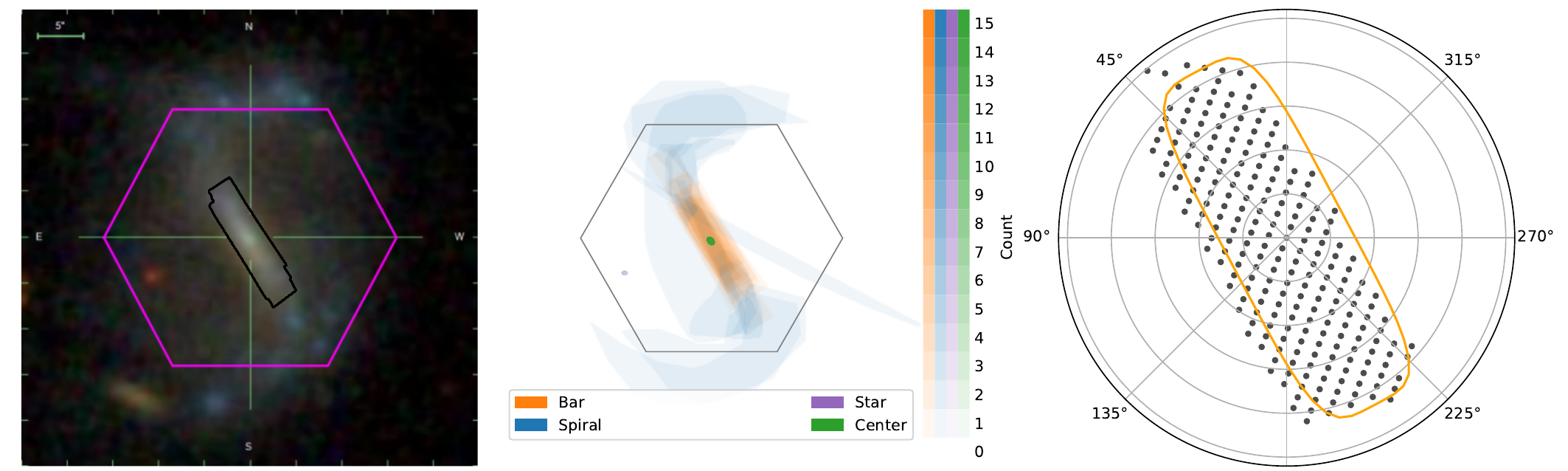}
        \caption{Figure demonstrating the bar identification and fitting. From left to right: (a) $r$-band SDSS image of the galaxy with \texttt{PLATEIFU=8075-12705} overlaid with the MaNGA hexagonal FoV and the bar mask inferred from the drawing in the next panel, (b) drawings from GZ:3D including bar, spiral, galaxy centre and foreground stars, colour-coded by the number count of superimposed drawings, and (c) pixel inside the bar mask as shown in the leftmost panel plotted in deprojected polar coordinates, also shown is the best-fit generalised ellipse, which is used to determine bar position angle and length. Note the small offset between the pixel mask and the fitted ellipse. This happens in a few cases in our sample and is most likely an observational effect in the drawings, while we force the ellipse representing the bar to be centred with respect to the galaxy centre.}
        \label{fig:zoo_lengths}
    \end{figure*}

    The ``average'' outline of a bar is obtained from the bar masks in GZ:3D. For the bar masks we choose a count threshold of $20\%$, in other words, a pixel is considered to be part of the bar if at least 3 out of 15 people included it within the outline. Choosing a threshold is a compromise between having a secure classification but a small bar or a looser classification and a large bar. In previous studies, a count fraction of $20\%$ has been proven to work reasonably well \citep{Fraser-McKelvie2019, Krishnarao2020}. We shall show later that our choice results in a good agreement with independent bar measurements.

    In principle it is possible to directly obtain the length and position angle of the bar from the bar masks by selecting - within the mask - the pixel that is farthest away from the galactic centre as reference point of the bar end \citep[see, e.g.,][]{Krishnarao2020}. Here, we choose a different path and fit the outer envelop of the mask with a generalised ellipse as defined in \citet{Athanassoula1990} in Cartesian coordinates (x,y):

    \begin{equation}
    \label{eq:gen_ell}
        \left( \frac{|x|}{a} \right)^c + \left( \frac{|y|}{b} \right)^c = 1.
    \end{equation}

    In the previous equation $a$ and $b$ are the semi-major and semi-minor axis of the ellipse, respectively, and $c$ is a boxyness parameter, with $c>2$ producing boxy shapes that approach a rectangle for increasing values. The fit is done in polar coordinates, where each bar pixel is plotted in the plane of the galaxy disc given its position angle $\Phi$ relative to the disc position angle and its deprojected galactocentric distance $r$. Replacing $x=r \cos{(\Theta)}$ and $y=r \sin{(\Theta)}$ in equation \ref{eq:gen_ell}, with $\Theta = \Phi - \Phi_{\rm ell}$ the position angle of each pixel relative to the angle of the semi-major axis of the ellipse, and re-shuffling leads to:

    \begin{equation}
    \label{eq:gen_ell_polar}
        r(\Phi) = \frac{a b}{\left(b^c |\cos{(\Phi - \Phi_{\rm ell})}|^c + a^c |\sin{(\Phi - \Phi_{\rm ell})}|^c\right)^{1/c}}
    \end{equation}

    Solving a least-square regression of equation \ref{eq:gen_ell_polar} for given sets of pixel coordinates ($r_{\rm pix}$,$\Phi_{\rm pix}$) results in bestfit parameters $a$, $b$, $c$ and $\Phi_{\rm ell}$ for the ellipse. The position angle and the bar length are then obtained as $PA_{\rm bar}=\Phi_{\rm ell}$ and $L_{\rm bar}=a$, with the constraints that $PA_{\rm bar}\in [0^\circ,180^\circ]$ and $a>b$. The procedure is exemplified in Fig. \ref{fig:zoo_lengths}. Panel a) shows an image of the galaxy overlaid with the bar mask, panel b) shows the mask drawings in GZ:3D, similar to figure 3 in \citet{Masters2021}, panel (c) shows the bar pixel mask in polar coordinates with the best-fitted ellipse.

    \subsubsection{Distribution of bar lengths}

\begin{figure}
	\centering
	\includegraphics[width=\columnwidth]{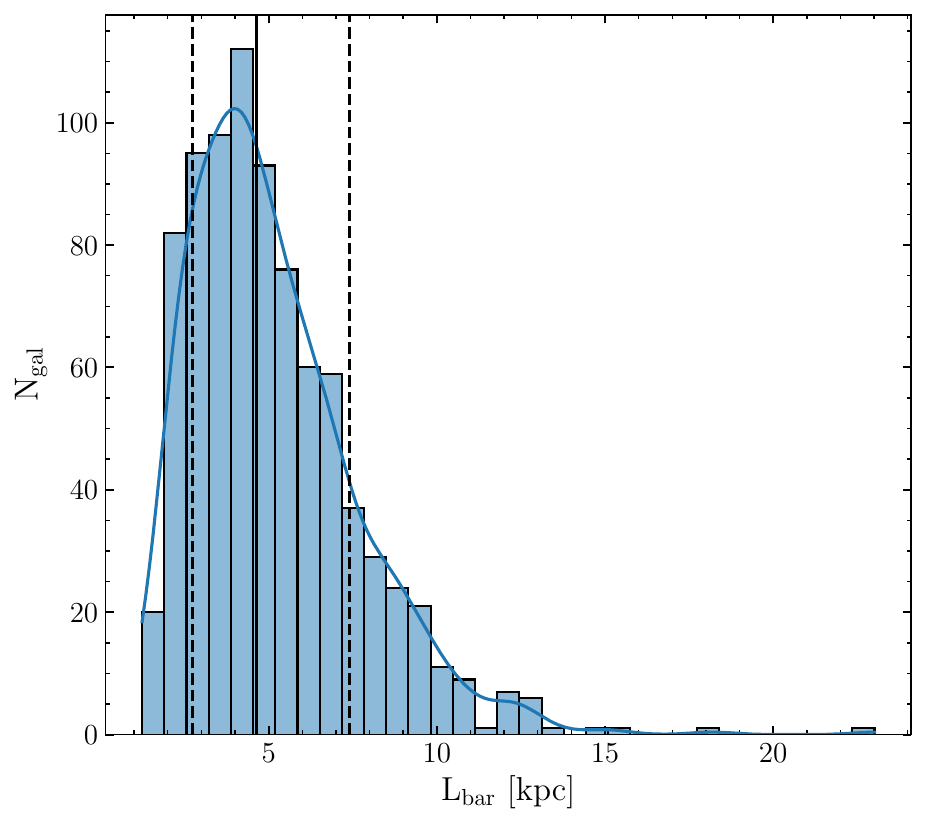}
        \caption{Distribution of physical bar lengths in the sample. The vertical solid and dashed lines mark the median and the 16th and 84th percentiles, respectively.}
        \label{fig:lengths}
    \end{figure}

    The overall distribution of bar lengths are presented in Fig. \ref{fig:lengths} and listed in Table \ref{tbl:bar_lengths}. The bars have lengths between 1 and 23$\,$kpc, with a median of ($4.6 \pm 0.1$)$\,$kpc and a sharp drop below 2$\,$kpc. The lower cutoff is likely associated with the resolution limit of the SDSS data ($\sim$0.5-2.5$\,$kpc with a median of 1.0$\,$kpc). The distribution is typical for strong bars in the local universe given the resolution limit \citep[e.g.][]{Barazza2008,Gadotti2009,Hoyle2011,Erwin2019,Geron2023} and is also in fair agreement with a resolution-matched analysis of barred galaxies in the cosmological simulation TNG50 \citep{Nelson2019,Pillepich2019} that reports $4.1 \pm 0.2\,$kpc \citep{Frankel2022}. Yet, while the distribution in TNG50 agrees with our measurements in MaNGA, \citet{Frankel2022} find larger bars in MaNGA. 

    \subsubsection{Comparison with literature}

        \begin{figure*}
            \centering
            \includegraphics[width=\linewidth]{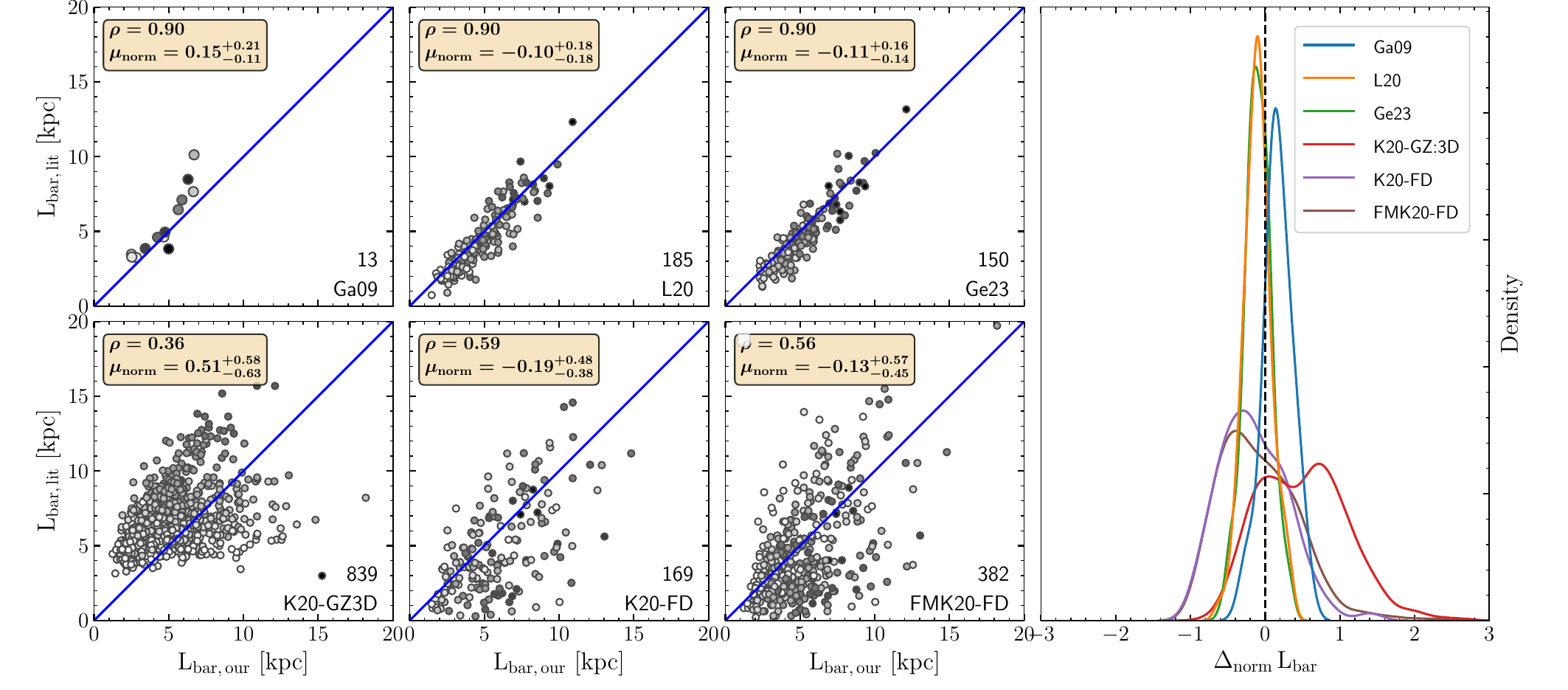}
            \caption{Comparison of bar lengths to values from the literature. The six squared panels on the left side show one-to-one comparisons to (from left to right, top to bottom) \citet{Gadotti2009}, \citet{Lin2020}, \citet{Geron2023}, \citet{Krishnarao2020} using GZ:3D and fourier decomposition, and \citet{FraserMcKelvie2020a}. The number of galaxies in common is listed in the bottom-right of each panel. The Pearson correlation coefficient and the normalised median difference is provided in the top-left. The greyscale shows the number of resolution elements per bar length as measured by us with darker shades representing better resolved bars. The right-most panel shows the distribution of the differences between the literature and our values.}
            \label{fig:lengths_comparisons}
        \end{figure*}

        In this paper, we put forward a novel approach for measuring bar lengths by combining crowd-sourced drawings with geometrical 2D fitting. Determining the extent of a bar and the definition of the length by itself are highly debated topics, see e.g. \citet{Ghosh2023} for a recent review and a comprehensive test and comparison of various methods. A few common techniques among observers include visual measurement \citep{Hoyle2011,Herrera-Endoqui2015,Geron2023}, 2D photometric decomposition \citep{Gadotti2009}, Fourier decomposition \citep{Aguerri2000,Laurikainen2006,Kraljic2012}, isophotal ellipse fitting \citep{Wozniak1991,Wozniak1995,Erwin2005c,Menendez-Delmestre2007} or a visually guided combination of these \citep{Herrera-Endoqui2015}.

        More recently, a dynamical measure of the length based on the extent of bar orbits has been proposed and tested on N-body models \citep{Petersen2023} and the Milky Way \citep{Lucey2023}

        A further complication seems to be the variable and asymmetric interaction between bars and spiral arms that leads to a fluctuation and underestimation of the bar length \citep{Hilmi2020}.

        Given that our sample is based on well-studied SDSS data, we are able to directly compare our measurements to several works in the literature with varying methods and sample overlap. In Fig. \ref{fig:lengths_comparisons}, we plot our results in comparison to lengths from photometric decomposition in \citet{Gadotti2009}, ellipse fitting in \citet{Lin2020}, visual measurements in \citet{Geron2023}, GZ:3D using the maximum pixel distance in \citet{Krishnarao2020}, and Fourier decomposition in \citet{Krishnarao2020} and in \citet{FraserMcKelvie2020a}. We find an excellent agreement with the first three of the aforementioned studies. The scatter is significantly larger in the latter three. Interestingly, bars are longer in \citet{Krishnarao2020} by 50\% using the same bar masks but defining the length by the maximum pixel distance. Furthermore, bars are shorter by 19\% and 13\% in  \citet{Krishnarao2020} and \citet{FraserMcKelvie2020a}, respectively, using Fourier decomposition.

        Evidently, the definition of bar length and method of measurement influence the outcome. In spite and because of these differences, it is indispensable to clearly define the method and keep it consistent throughout the analysis.
    
    \subsection{Stacking procedure}
    \label{sect:analysis_stacking}

        \begin{figure*}
            \centering
            \includegraphics[width=\linewidth]{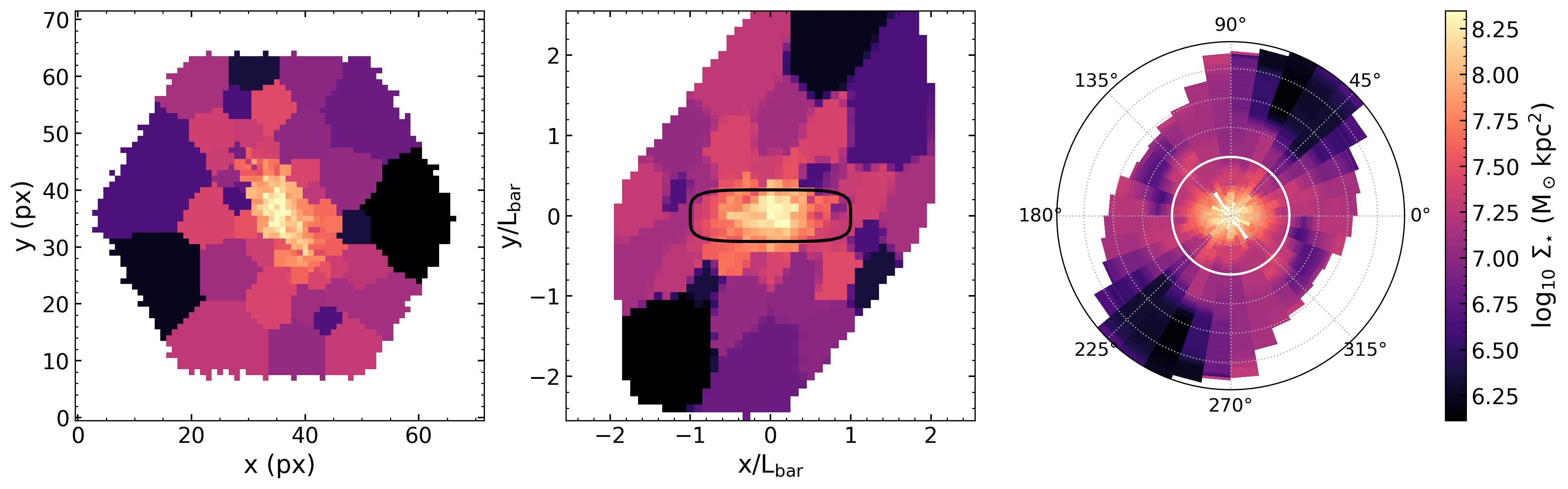}
            \caption{Demonstration of stellar population map manipulation for stacking procedure. Shown is the stellar mass surface density of galaxy \texttt{PLATEIFU=8075-12705}. \emph{Left:} original Voronoi-binned map from the \texttt{FIREFLY} catalogue. \emph{Middle:} map rotated to align bar to x-axis and scaled to bar length (map is interpolated in this figure for illustration purposes only), overplotted with fitted bar shape. \emph{Right:} map is deprojected to inclination zero and regridded onto polar coordinates, the white circle indicates the bar radius.}
            \label{fig:proc}
        \end{figure*}

        With the lengths and position angles of the bars at hand, we are now able to project our measurements onto a common coordinate grid, which we define as follows: We use polar coordinates ($R$, $\theta$) with $R=1$ at the bar radius and $\theta=0\degr$ at the bar position angle, i.e. the bar is along the x-axis and the radius is normalised to the length of the bar. Azimuthal bins are of $10\degr$ width. Radial bins are of $\rm 0.2\,L_{bar}$ width with a $\rm 0.04\,L_{bar}$ step size to calculate `running averages'. This choice was made after testing different setups and it was identified to be the best compromise between having a well-sampled median and tracing small scale radial fluctuations. 

        To sample the distribution of stellar populations in each galaxy, we take pixel values at their discrete centre positions. Pixel coordinates are transformed into polar coordinates and face-on galaxy plane projection. Afterwards, radii are normalised to the bar length and rotated, such that the bar is on the x-axis, as described above. This is illustrated in Fig. \ref{fig:proc}. The final preparatory step before stacking is azimuthal normalisation of the respective stellar population parameter. In steps of $\rm 0.04\,L_{bar}$ we calculate the median age, total metallicity [Z/H] and $\Sigma_\star$ within a ring of width $\rm 0.2\,L_{bar}$ around the $\rm 0.04\,L_{bar}$-ring and subtract that median from each single pixel value of the narrower $\rm 0.04\,L_{bar}$-wide ring. Each pixel value is now azimuthally normalised and transformed onto the new common polar grid but all values are so far only collected within their polar-coordinate bins and not yet averaged or stacked.

\begin{figure}
	\centering
	\includegraphics[width=\columnwidth]{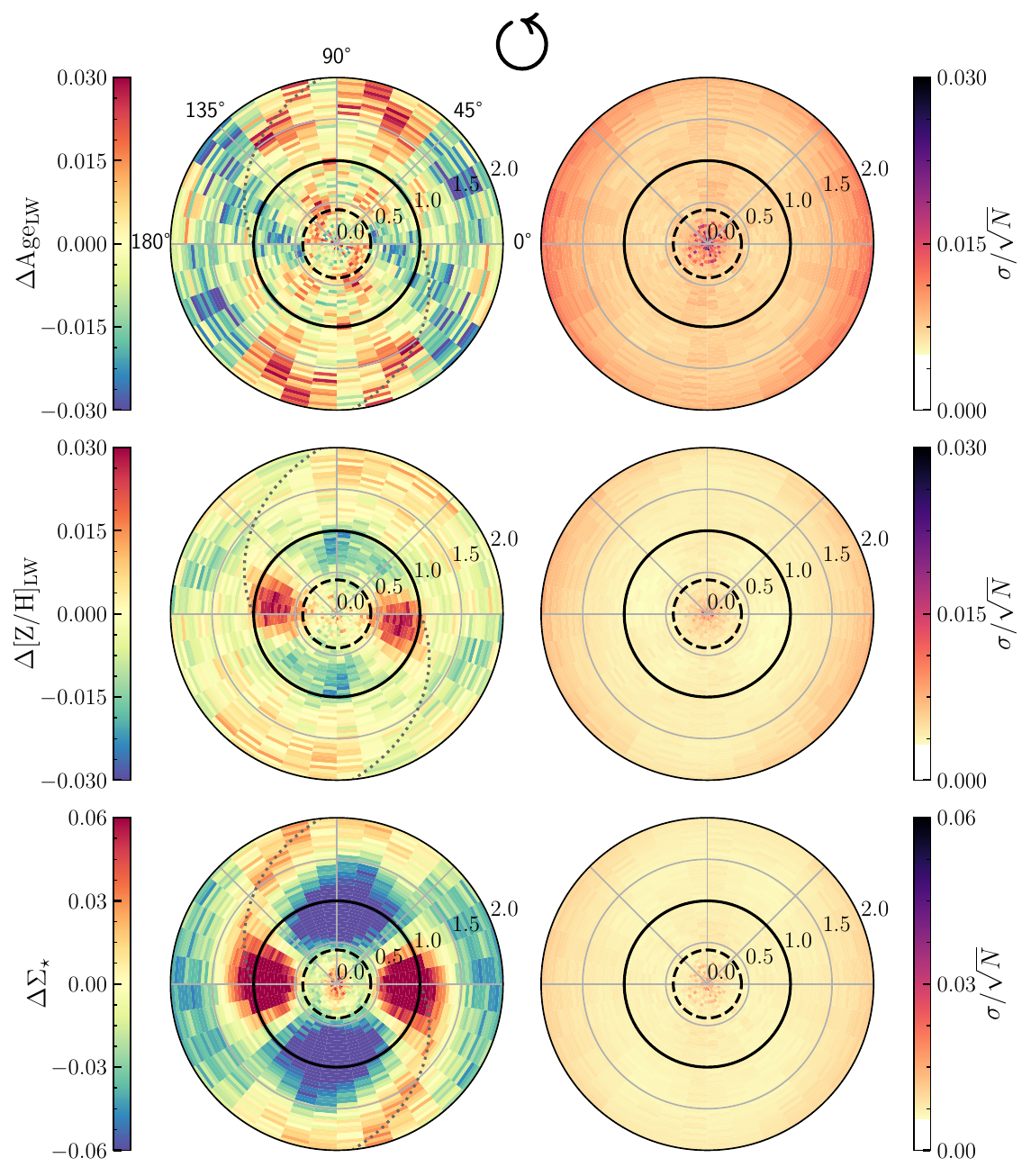}
        \caption{Azimuthal variations of stellar populations for all high-mass galaxies with determinable sense of rotation, showing, from top to bottom, light-weighted age, [Z/H] and $\Sigma_\star$. Left panels show medians per polar grid cell, while the right panels show significance maps of these medians. The black solid circle marks the bar radius and the dashed circle marks the average 2$\times$FWHM of the PSF across the subsample. The dotted line follows a log-spiral function with a pitch angle of $~25.2\degr$. Zero degree is towards the right, 90 degree towards the top. The arrow indicates the anti-clockwise rotation of the galaxies.}
        \label{fig:polar_example}
    \end{figure}

        With the data in the new layout from the previous steps, we can now take any combination of galaxies and stack their data. For each cell on the polar grid, we then calculate the median to produce the final stacked polar stellar population maps. During this process, in order to improve statistics, we assume point symmetry in each galaxy disc, specifically, we rotate the third and fourth quadrant by 180$\degr$ and average them with the first and second quadrant, respectively. In the case of a rotating bar, for example, this can be thought of as averaging leading edges of the bar with leading edges and trailing edges with trailing edges. However, for this to work properly, a knowledge of the sense of rotation is necessary, which we obtain from visually inspecting all galaxies and assuming that spiral arms are trailing. This is only possible for about 2/3 of our galaxies, as some galaxies do not have visible spiral arms. In Fig. \ref{fig:polar_example}, in the next section, we restrict the sample to only those galaxies that we know the rotation of. For all further analysis, we include all galaxies and only rotate galaxies with visible spiral arms.

\section{Results}
\label{sect:results}

    \begin{figure*}
        \centering
        \includegraphics[width=\linewidth]{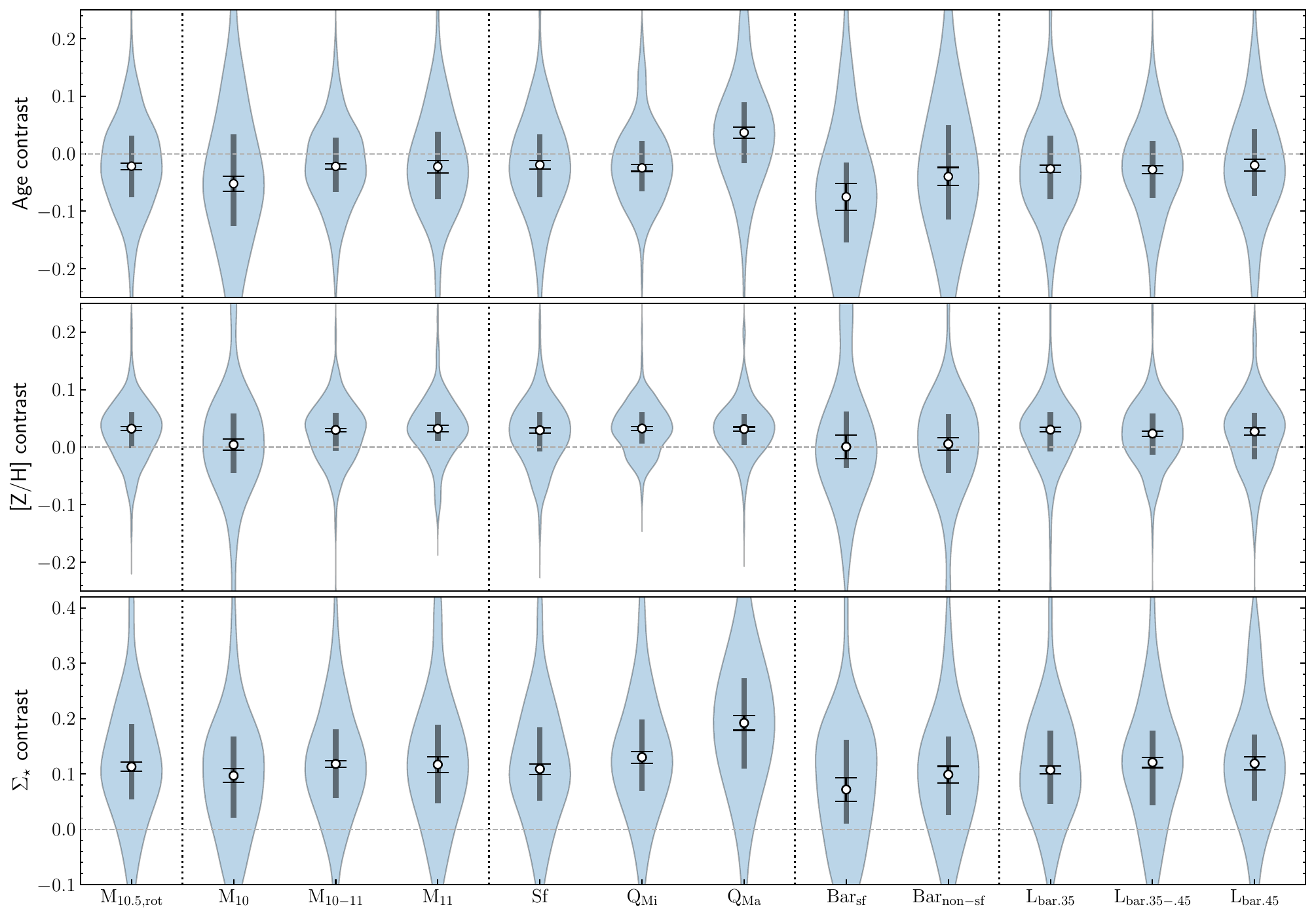}
        \caption{Bar-to-inter-bar contrast of age, [Z/H] and $\Sigma_\star$ for all subsamples discussed in the text. Violins show the distribution of individual galaxies. Marked are the medians (white dots), interquartile ranges (vertical grey bars) and standard errors on the medians (black error bars).}
        \label{fig:violins}
    \end{figure*}

    In this section, we present 2D polar maps of the azimuthal differences of light-weighted mean stellar ages, stellar metallicities and stellar mass surface densities. In addition, we define the average region on these plots that are within the bar and a comparison-region perpendicular to it in the inter-bar region and calculate the contrast in age, metallicity and stellar mass surface density between bar and inter-bar.

    \subsection{General distributions}
    \label{sect:results_general}

        \subsubsection{2D polar maps}

            In Fig. \ref{fig:polar_example} we show the azimuthal difference plots for all massive ($M_\star > 10^{10.5}\,M_\odot)$ galaxies with known sense of rotation (number of galaxies $N_{\rm gal}=374$). These plots do not show absolute values, radial gradients are implicitly removed by our analysis. For illustration purposes, we mirror and flip the upper half of the polar plot [0$\degr$,180$\degr$] onto the lower half [180$\degr$,360$\degr$]. Galaxies rotate counter-clockwise in this representation and we remind that bars are elongated along the x-axis within the circle with $R=1$. We will first qualitatively discuss our observations and we will provide a quantitative comparison in the next subsection.

            The figure shows a strong difference of stellar mass surface densities between the bar and inter-bar region. The difference is strongest between 0.5 and 1.2$\times L_{\rm bar}$. It is not surprising that bars appear here as stellar over-densities, because during the formation and evolution of a bar, most stars within the bar radius get trapped on elongated x1 orbits forming the backbone of the bar \citep{Contopoulos1980,Athanassoula1983,Pfenniger1984,Skokos2002,Skokos2002a}. Our results are qualitatively consistent with \citet{Krishnarao2022}. In fact, these authors used the azimuthal difference of stellar densities to find the location of `dark gaps', i.e. low surface brightness areas in the inter-bar region, which, as they showed, coincides with the inner ultra harmonic (4:1) resonance of a barred galaxy (see also \citealp{Kim2016, Buta2017}). In addition, in our $\Delta \Sigma_\star$ map we see a faint over-density signature of spiral arms, connecting to the bar ends and winding outwards. We overplotted on the figure a simple log-spiral function starting at ($r=1$,$\Phi=180\degr$) with a pitch angle of $\alpha=25.2\degr$ that seems to follow the $\Delta \Sigma$ feature nicely. The appearance of this spiral is not necessarily expected, as only the bars are aligned by construction. The spiral arm signature could be interpreted as average or dominant spiral arm with a preferred pitch angle. Finally, we point out that the central $\rm \sim 0.5\,L_{bar}$ shows very small to zero differences along the azimuth, which is a recurring feature in all of the following figures of this kind in this work.\footnote{We note that lower significance of the average derived parameters in the centre are driven by smaller numbers $N$ of data points per polar grid cell.} The absence of any azimuthal variations in the centre can easily be explained by a combination of factors. Firstly, beam smearing due to limited spatial resolution will smooth out some of the potential differences. The average resolution is shown in each figure as dashed circle. Secondly, central morphological structures such as hot spheroids or cold nuclear discs are to first degree axisymmetric structures, with little azimuthal variations, that might dominate the stellar populations in the central region. The size of nuclear discs is usually 5-10\% of the size of the bar \citep{Gadotti2020}, hot spheroids can have a wider range, yet it is still disputed if significant spheroids exist in barred galaxies \citep{Gadotti2020}. Thirdly and finally, box-peanut structures, which are the central inherent parts of bars that buckle out of the disc plane, are wider than the width of the long part of the bar and may wash out central azimuthal differences. They are suggested to be present in almost all massive barred galaxies \citep{Erwin2023}.

            Similar to stellar density, also stellar metallicity has an above-average value along the bar. This metallicity excess is, in contrast to $\Delta \Sigma_\star$, contained within the bar radius. Additionally, we observe a small asymmetry between the leading and the trailing edge of the bar\footnote{Given the counter-clockwise rotation of the galaxies in this figure, the leading edge of the rotating bar, can be identified approximately between $0\degr$-$30\degr$ and $180\degr$-$210\degr$, while the trailing edge is at $330\degr$-$360\degr$ and $150\degr$-$180\degr$}: The metal excess extends azimuthally further away from the bar on the trailing edge (10$\degr$-20$\degr$).

            The stellar age map does not display any clear trend. The scatter across the map of the medians is as noticeable as the slightly lower significances compared to that of metallicity and stellar mass density. There is a slight preference for younger ages along the trailing edge of the bar, which continues further outward beyond the bar radius at larger distances from the bar major axis. 

            Back to metallicity and mass density differences, we would like to point out a very subtle feature seen in Fig. \ref{fig:polar_example}, as well as in star-forming galaxies in Fig. \ref{fig:trends_sfr}. In both subsamples, spiral arm features are visible in the mass densities. In the metallicity maps, we see a very subtle trend, where regions leading the spiral arm are more metal-poor and regions trailing the spiral arm are more metal-rich. If real, this points towards radial migration. When radial migration driven by spiral arms \citep{Sellwood2002} and enhanced by bars \citep{Minchev2010} happens, metal-rich stars migrate outwards behind the spiral arm and metal-poor stars migrate inwards in front of spiral arms \citep[e,g,][]{Grand2016}. Alternatively or additionally, azimuthal metallicity variations can also arise naturally without invoking radial migration by different responses of kinematically hot and kinematically cold stellar populations to spiral density perturbations \citep[e.g.][]{Khoperskov2018a}. At this point, we remain cautious with this observation as a proper analysis with clear identification and alignment of spiral arm regions is needed to confirm this hypothesis. If real, however, it is important because it is detected in integrated stellar light in external galaxies.

        \subsubsection{Bar versus inter-bar}

            To provide more quantitative measurements of the differences between the bar and the inter-bar, based on the $\Delta\Sigma_\star$ map, we define an average region in the polar plots that is dominated by the bar to be within [-20\degr,20\degr] and [160\degr,200\degr] in azimuth and within [0.5,1.0] in normalised radius. The comparison region in the inter-bar region is defined to be within [70\degr,110\degr] and [250\degr,290\degr] at the same radial range [0.5,1.0]. The age contrast is then defined as $\langle \Delta \rm Age\rangle_{bar}-\langle\Delta Age\rangle_{inter-bar}$, and equally for metallicity and mass surface density contrast. Here, the average is taken over the bar (inter-bar) region but for individual galaxies. The distribution of the bar-inter-bar contrasts is shown in Fig. \ref{fig:violins} and values are tabulated in Table \ref{tbl:contrasts_miles}.\footnote{If instead we calculate the average contrasts of the whole sample directly from the stacked maps, they agree with the median contrasts (within the error bars) from the distribution of the individual-galaxy calculated values.} 

            The contrast distributions are shown for different subsamples along the x-axis, with the massive galaxies with known rotation, discussed in the previous subsection, on the very left. In this subsample, bars have on average higher stellar mass surface densitities by $0.113\pm0.009$\,dex, higher metallicity by $0.032\pm0.004$\,dex and are younger by $-0.022\pm0.006$\,dex. Quoted are the medians and the error on the median estimated as $\sqrt{\pi/2 \cdot \sigma^2/N}$, where $\sigma$ is the standard deviation of the mean and $N$ is the number of galaxies. Looking at the full distributions and the inter-quartile ranges of this subsample, there is a large spread between negative and positive contrasts of stellar ages with 40\% of bars having actually older stellar populations. The distribution of stellar metallicities and mass surface densities tell a clearer story.

            In the following, we will discuss how the azimuthal stellar population variations change across different subsamples of galaxies probing ranges in mass, distance from SFMS, star formation in the bar itself, bar length, as well as different stellar population libraries for modelling the spectra.
    
    \begin{table*}
        \renewcommand{\arraystretch}{1.3}
	\caption{Bar to inter-bar contrast for different subsamples of galaxies}
	\label{tbl:contrasts_miles}
	\centering
        \resizebox{\textwidth}{!}{%
	\begin{tabular}{rrrrrrrr} 
		\hline
			& $\rm M_{10.5,rot}$ & $\rm M_{10}$ & $\rm M_{10-11}$ & $\rm M_{11}$ & $\rm Sf$ & $\rm Q_{Mi}$ & $\rm Q_{Ma}$ \\
		\hline
		Age contrast &
            $-0.022^{+0.053}_{-0.055}$ & 
            $-0.052^{+0.086}_{-0.074}$ & 
            $-0.022^{+0.050}_{-0.044}$ & 
            $-0.022^{+0.061}_{-0.057}$ & 
            $-0.019^{+0.053}_{-0.056}$ & 
            $-0.025^{+0.047}_{-0.040}$ & 
            $0.037^{+0.053}_{-0.053}$ \\ 
            $[Z/H]$ contrast &           
            $0.032^{+0.028}_{-0.034}$ & 
            $0.004^{+0.055}_{-0.049}$ & 
            $0.030^{+0.030}_{-0.036}$ & 
            $0.032^{+0.029}_{-0.021}$ & 
            $0.029^{+0.031}_{-0.037}$ & 
            $0.032^{+0.028}_{-0.027}$ & 
            $0.032^{+0.026}_{-0.028}$ \\
            $\Sigma_\star$ contrast & 
            $0.113^{+0.077}_{-0.059}$ & 
            $0.097^{+0.071}_{-0.076}$ & 
            $0.118^{+0.062}_{-0.062}$ & 
            $0.117^{+0.072}_{-0.070}$ & 
            $0.109^{+0.076}_{-0.058}$ & 
            $0.130^{+0.069}_{-0.061}$ & 
            $0.192^{+0.082}_{-0.082}$ \\
		\hline
            & $\rm Bar_{sf}$ & $\rm Bar_{non-sf}$ & $\rm L_{bar.35}$ & $\rm L_{bar.35-45}$ & $\rm L_{bar.45}$  & $\rm All_{Mi}$ & $\rm All_{Ma}$ \\
            \hline
            Age contrast & 
            $-0.075^{+0.059}_{-0.079}$ & 
            $-0.040^{+0.089}_{-0.074}$ &
            $-0.026^{+0.058}_{-0.053}$ & 
            $-0.028^{+0.050}_{-0.049}$ & 
            $-0.020^{+0.063}_{-0.054}$ & 
            $-0.026^{+0.057}_{-0.051}$ &
            $0.021^{+0.068}_{-0.076}$ \\ 

            $[Z/H]$ contrast & 
            $0.001^{+0.061}_{-0.037}$ & 
            $0.006^{+0.052}_{-0.051}$ & 
            $0.030^{+0.030}_{-0.038}$ & 
            $0.024^{+0.035}_{-0.037}$ & 
            $0.027^{+0.033}_{-0.048}$ & 
            $0.028^{+0.033}_{-0.040}$ & 
            $0.025^{+0.034}_{-0.037}$ \\   
            $\Sigma_\star$ contrast & 
            $0.072^{+0.090}_{-0.062}$ & 
            $0.099^{+0.069}_{-0.073}$ &
            $0.107^{+0.071}_{-0.061}$ & 
            $0.121^{+0.058}_{-0.077}$ & 
            $0.119^{+0.053}_{-0.067}$ & 
            $0.113^{+0.065}_{-0.067}$ & 
            $0.157^{+0.085}_{-0.085}$ \\ 
            \hline
	\end{tabular}}
	\medskip\\
	\raggedright
        Notes: Quoted are the medians with the upper and lower limit representing the difference to the 75th and 25th percentiles of the distribution
    \end{table*}	    
 
    \subsection{Trend with galaxy mass}

        \begin{figure*}
            \centering
            \includegraphics[width=\linewidth]{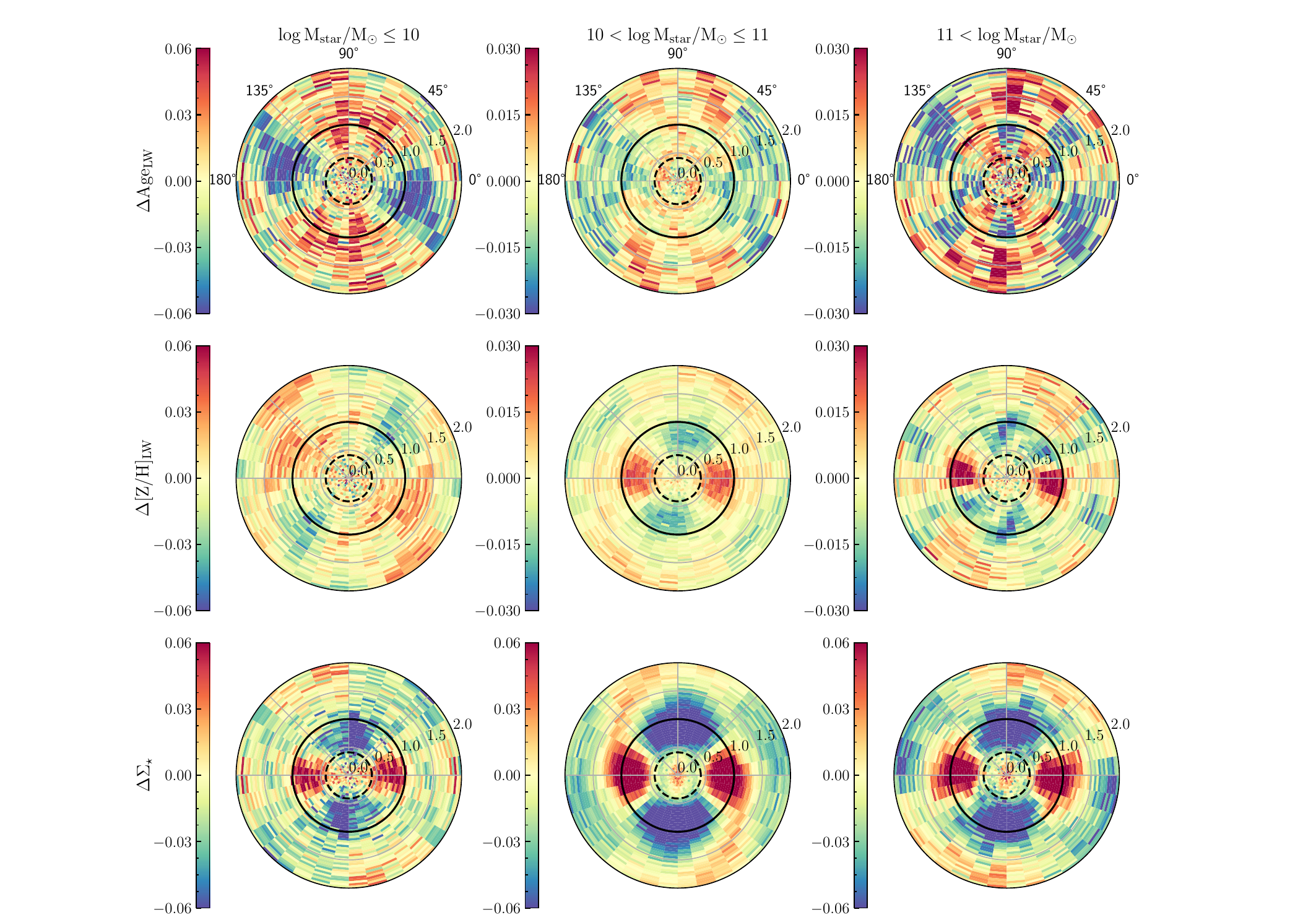}
            \caption{Trends with galaxy stellar mass, showing three stellar mass bins from low-mass to high-mass from left to right. Figure format is the same as in the left column of Fig. \ref{fig:polar_example}. Here we show only the left columns, i.e. the medians, for clarity.}
            \label{fig:trends_mass}
        \end{figure*}

        We first investigate how the previously described features in the azimuthal variation maps depend on total stellar mass. We divide our sample in three mass bins, with $\log_{10} (M_\star/M_\odot) <10$, between 10 and 11 and, >11, this includes $N_{\rm gal}=160$, $N_{\rm gal}=550$ and $N_{\rm gal}=135$ galaxies, respectively. To increase statistics, we include all galaxies whether or not we know the sense of rotation. Galaxies with known rotation, are positioned to rotate counter-clockwise, all others are rotating randomly. The results are presented in Fig. \ref{fig:trends_mass} and Fig. \ref{fig:violins}, columns 2-4. For simplicity, we do not show the significance maps in the polar plot, which are for these results of similar amplitude as in Fig. \ref{fig:polar_example}.

        All subsamples of galaxies presented here show higher $\Sigma_\star$ and higher [Z/H] along the bar, while age is less conclusive with slightly younger ages, but note again the large overlap to older ages for $>33$\% of all subsamples. In addition, there is a clear increase of the excess of stellar mass density and metallicity in the bar for galaxy masses beyond $\log_{10} (M_\star/M_\odot) =10$. The lowest-mass galaxies have the least massive bars (0.097\,dex) with only marginally increased metallicity (0.004\,dex) and the youngest population ($-0.052$\,dex). The azimuthal angle of the metallicity peak in the lowest-mass galaxies is offset compared to the higher mass galaxies and not clearly located within the bar.

    \subsection{Trend with distance from SFMS}

        \begin{figure*}
            \centering
            \includegraphics[width=\linewidth]{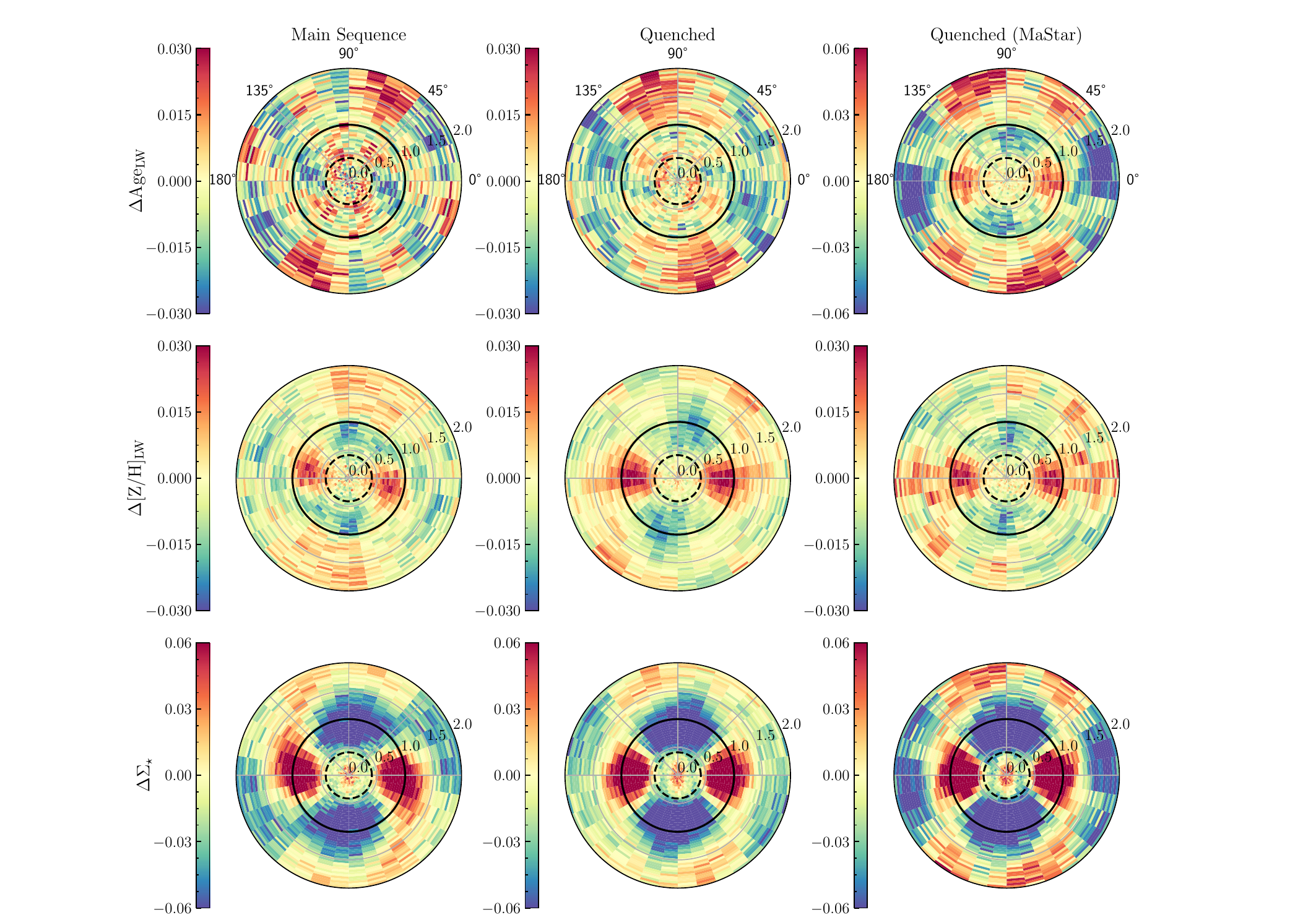}
            \caption{Trends with distance from SFMS from left to middle panel and stellar population model library from middle to right panel. Figure format same as in Fig. \ref{fig:polar_example}.}
            \label{fig:trends_sfr}
        \end{figure*}

        Further, we probe the effect of ongoing star formation. On the one hand, one might imagine that current star formation will imprint the gas metallicity content of today's galaxy on the average stellar population, whose light-weighted average metallicity will be dominated by younger populations, while at the same time quenched galaxies reflect the metal content of a chemical state earlier in cosmic time. Quiescent galaxies might also have gone through more violent quenching mechanisms, processes which can affect the metal distribution in galaxies. On the other hand, a quiescent barred galaxy might also imply a more mature bar that had more time to leave its imprint on the host galaxy.

        In Fig. \ref{fig:trends_sfr} left and middle columns, and Fig. \ref{fig:violins} columns 5-6, we compare galaxies on the star-forming main sequence (SFMS, $N_{\rm gal}=250$), as defined by the relation in \citet[][see also Fig. \ref{fig:sample}]{Renzini2015} to quenched galaxies ($N_{\rm gal}=235$), which we define to be all galaxies at least 0.5$\,$dex below the SFMS. We limit the selection to galaxies with $M_\star>10^{10.5}\,M_\odot$, to probe a comparable mass distribution in both subsamples. 

        We observe that quenched galaxies have a higher mass surface density contrast (0.130\,dex compared to 0.109\,dex) and a slightly higher metal excess along the bar (0.032\,dex compared to 0.029\,dex) and even a little beyond. We feel confident that the excess beyond the bar radius is not due to a systematic bias in the bar lengths -- e.g. biased length estimation due to higher dust obscuration in star-forming galaxies -- given the similarity of the stellar mass distribution. Stellar populations are on average younger in the bar in both subsamples, but there is again a large overlap with a significant fraction of galaxies with older populations in the bar (40\% and 36\%). Beyond the bar radius, we observe a tendency to younger ages along the extension of the bar major axis in quenched galaxies. We will discuss the comparison to the quenched galaxies fitted with MaStar stellar population models, as seen in the right column of these plots in Sect. \ref{Sect:trendSSP} and Appendix \ref{apx:ssp}.
        


    \subsection{Trend with SF inside the bar}
    \label{sect:trendSFbar}

        \begin{figure*}
            \centering
            \includegraphics[width=\linewidth]{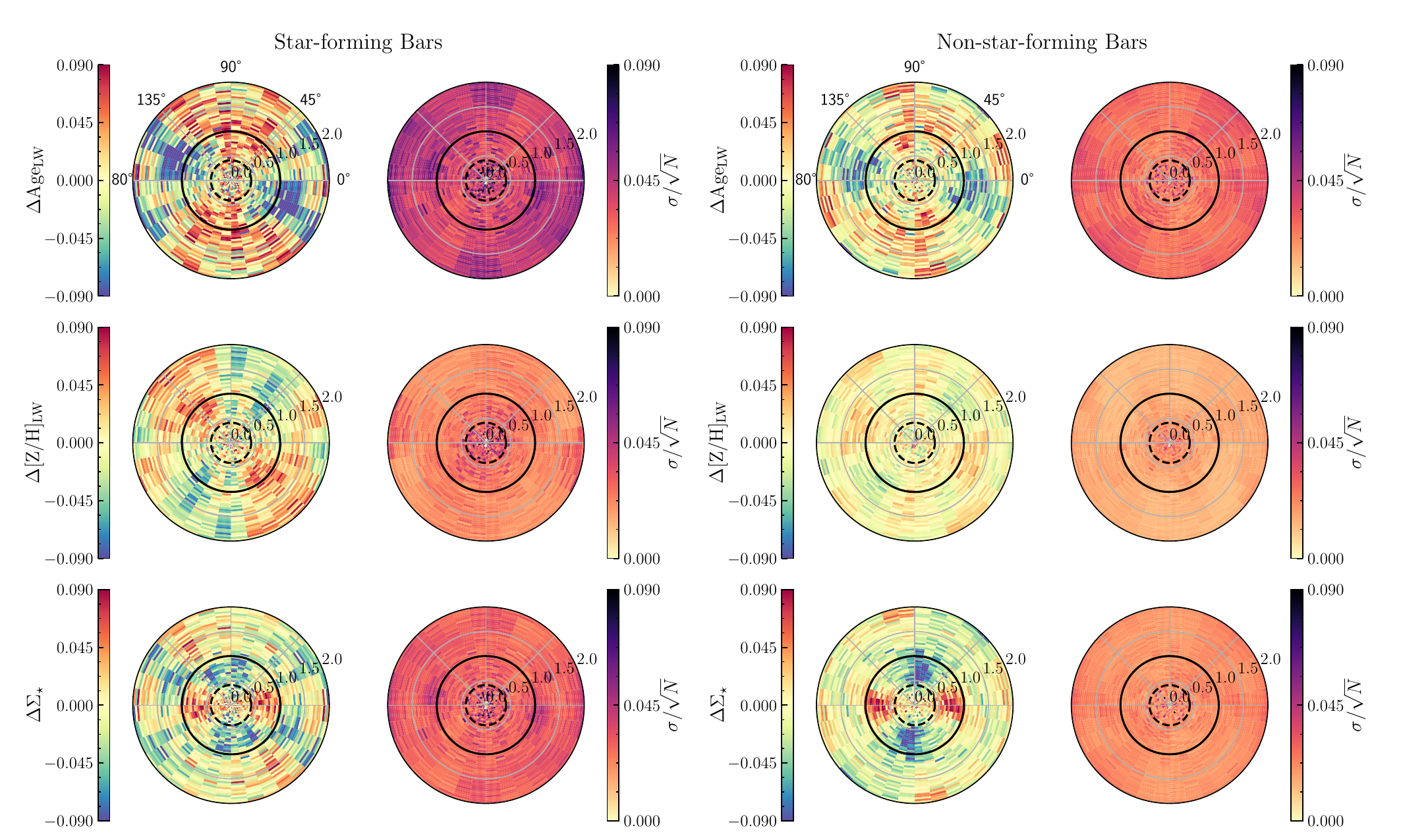}
            \caption{Trends with star formation in the bar. Comparing star-forming bars (left) to non-star-forming bars (right) of low-mass galaxies ($\log M_\star / M_\odot <10$). Figure format is the same as in Fig. \ref{fig:polar_example}. Significance maps are shown as the intrinsic scatter is larger than in the previous subsamples.}
            \label{fig:trends_sfbar}
        \end{figure*}

        Next, we compare the azimuthal variations between barred galaxies with star formation within the bar itself ($N_{\rm gal}=44$) to barred galaxies with no star formation in the bar ($N_{\rm gal}=116$). The classification is obtained for a subsample of galaxies from \citet{FraserMcKelvie2020a}, see our Fig. \ref{fig:sample} and Sect. \ref{sect:subsamples}. The comparison is shown in Fig. \ref{fig:trends_sfbar} and Fig. \ref{fig:violins} columns 8-9. We note that in this case we opt to show the significance maps for each panel to emphasise that the scatter in these measurements is significantly larger, which is mostly driven by the smaller number statistics and the typical smaller signal in low-mass galaxies.

        Bars are clearly recognisable by their stellar mass density excess along the major axis, as previously (0.072\,dex and 0.099\,dex), but with the lowest contrast in star-forming bars. In addition, on the one hand, star-forming bars have the youngest stellar population contrast ($-0.075$\,dex) compared to all other analysed subsamples. At the same time they have no clear metallicity enhancement along the bar (0.001\,dex), but they do show higher metallicities skewed by $30\degr$-$40\degr$ towards the trailing edge.
        
        On the other hand, non-star-forming low-mass bars show a marginal excess of metallicity (0.006\,dex) with on average younger populations ($-0.040$\,dex), but less strong than star-forming bars, and again with a large overlap to older ages.
        


    \subsection{Trend with bar length}

        \begin{figure*}
            \centering
            \includegraphics[width=\linewidth]{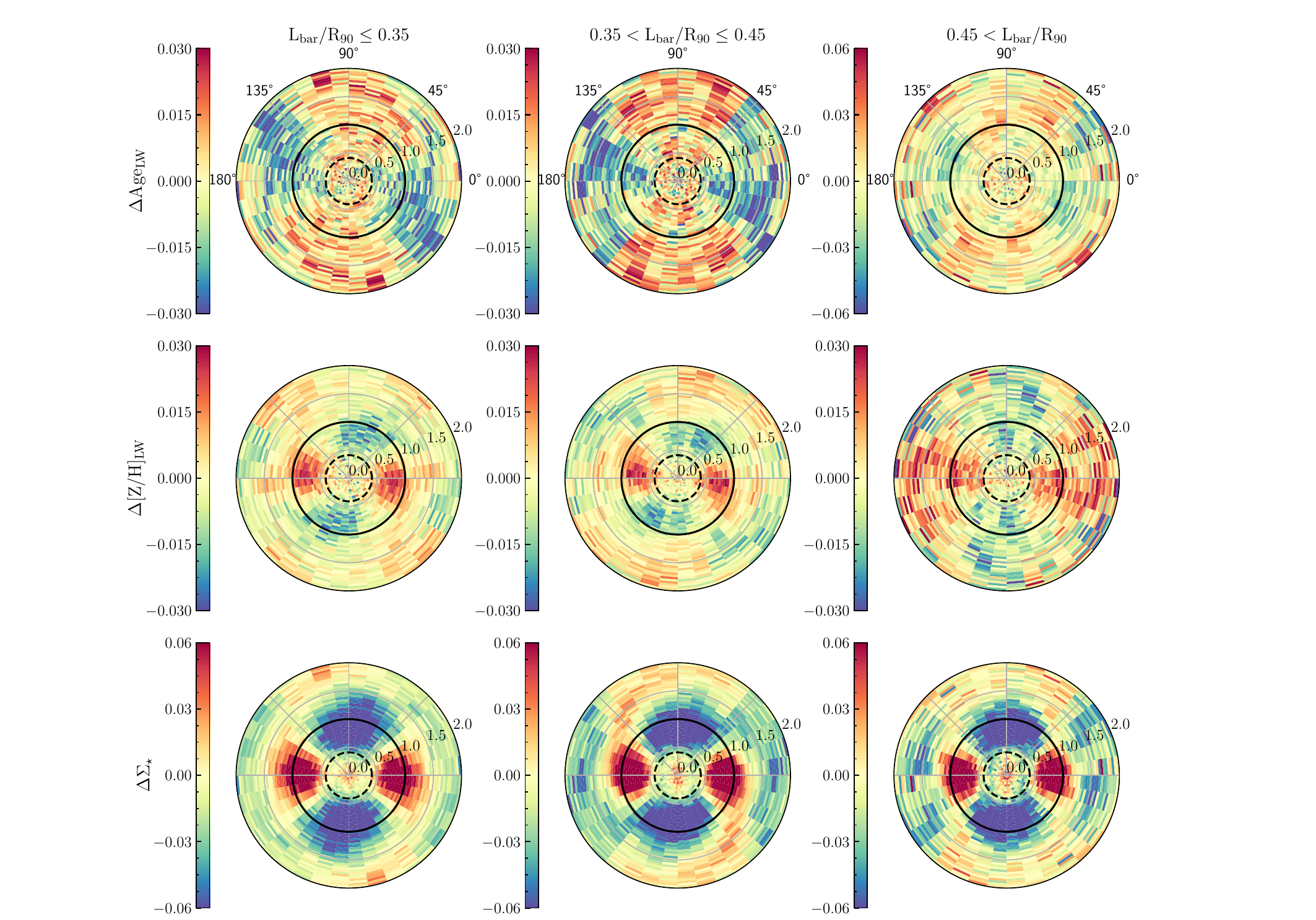}
            \caption{Trends with bar length from short to long bars from left to right. Figure format is the same as in Fig. \ref{fig:polar_example}.}
            \label{fig:trends_lbar}
        \end{figure*}

        The length of a bar scaled to the size of its host disc is potentially related to the evolutionary state of the bar. Some simulations indicate that bars grow over time  \citep{Tremaine1984b,Athanassoula2003,Martinez-Valpuesta2006}. Yet, other simulations show that the oldest bars form long and do not evolve much with time, while only bars that form at later times in the Universe form short and then tend to grow with time \citep{Fragkoudi2024}.

        We test if the length of the bar with respect to the disc has an impact on the distribution of stellar populations in Fig. \ref{fig:trends_lbar} and Fig. \ref{fig:violins} columns 10-12. We see very little difference between small bars ($L_\mathrm{bar}/R_{90} < 0.35$, $N_{\rm gal}=379$) and medium-size bars ($0.35<L_\mathrm{bar}/R_{90}<0.45$, $N_{\rm gal}=239$), with bars being massive, metal-rich and younger than their surroundings. However, the picture changes for long bars ($L_\mathrm{bar}/R_{90}>0.45$, $N_{\rm gal}=173$). Galaxies with such long bars show less of an azimuthal age difference. While the median age contrast is still $-0.020$\,dex, 37\% of the sample has older populations in the bar than in the inter-bar. These bars are still more metal-rich, and this trend now continues along the bar major axis out to twice the bar radius, the farthest distance probed in this analysis. This excess of metal-rich populations beyond the bar is similar to and partially even stronger than the observed trend in quenched galaxies. A correlation between both trends with quenching state and scaled bar length is in good agreement with long bars being more frequent in quenched galaxies, as found in \citet{FraserMcKelvie2020a} and also our Fig. \ref{fig:sample}.
        

    
    \subsection{Trend with SSP library} 
    \label{Sect:trendSSP}

        Most of the results presented in this paper are based on the stellar population fitting employing the MILES SSP library. This choice was based both on the observation that trends in the polar maps are seen more clearly and with less scatter, as well as for comparability to the literature given that MILES-based stellar population libraries have been frequently used over the last two decades. However, none of our conclusions throughout the paper changes if we were to present results stemming from the catalogue using MaStar SSPs, with the exception of stellar ages as discussed shortly in the following and with more detail in Appendix \ref{apx:ssp}.

        So far, it has become clear that azimuthal changes of stellar ages make the widest distribution across galaxies. While on average, stellar bars tend to be younger when fitted with MILES, there is always a significant overlap to bars with older ages. The few trends across subsamples of galaxies that have been observed in MILES are also consistent with the results from MaStar. Yet, it is important to highlight that stellar populations when fitted with MaStar show in general older average ages along the bar (or younger average ages perpendicular to it) than in the MILES catalogue. This is most noticeable in quenched galaxies and illustrated in Fig. \ref{fig:trends_sfr} right column. In this case, stellar populations are older in the bar region as compared to the inter-bar region at same radii. In contrast, using MILES, they are slightly younger. At this point the origin of this discrepancy remains unclear. Systematic differences between both \ffvac{} variants are not easy to track down. Standing out is the lack of very young templates in the MILES library and the longer wavelength extension allowed by MaStar-based SSP models. This has an impact on fitting young stellar population regions. In the case of quenched galaxies in Fig. \ref{fig:trends_sfr}, this seems to be a less important issue, though. We elaborate further on the differences when using these two libraries in Appendix \ref{apx:ssp}.
        

\section{Discussion}
\label{sect:discussion}

    \subsection{Metal-rich bars}

        The robust result throughout all explored subsamples is the metal excess of stellar populations in bars compared to the inter-bar regions in the inner disc. In other words, bars are metal-rich structures in disc galaxies. This result is consistent with our previous indicative finding in \citet{Neumann2020} based on a small sample of nine galaxies. Our new statistically comprehensive analysis presented in this paper puts the \citet{Neumann2020} result on firm grounds. Thereby, we are able to provide an extra-galactic reference for simulations and Milky Way and Local Group studies. 

        In fact, the Milky Way bar has been found to be more metal-rich than the disc  \citep{Wegg2019,Lian2021,Queiroz2021}, but opposite results have also been put forward \citep{Bovy2019,Eilers2022}. Additionally, resolved stellar population studies of M31, also suggest a metal-rich bar in the Andromeda galaxy \citep{Saglia2018,Gajda2021,Gibson2023}. Our results are also in good agreement with cosmological and N-body simulations \citep{Buck2018,Fragkoudi2020,Filion2023}. Finally, even nuclear (or inner) bars, those that are formed within the nuclear disc of barred galaxies, with sizes between 0.17 and 2.4$\,$kpc \citep{deLorenzoCaceres2020}, have been observed to be more metal-rich than the immediate surroundings \citep{Lorenzo-Caceres2013,Lorenzo-Caceres2019a,Bittner2021}.

        Several processes may contribute to explain the chemical difference of stellar population inside the bar as compared to the inter-bar region at the same radii. The true cause is likely to be a combination of some or all of the following scenarios:

        (i) N-body simulations have shown that during the formation of a bar stellar populations can get separated based on their kinematic properties. Young, metal-rich and kinematically cold stars get trapped on colder, more elongated orbits, forming a thinner part of the bar. Old, metal-poor and kinematically hot stars get trapped on hotter, rounder orbits, forming a thicker part of the bar \citep{Athanassoula2017,Debattista2017,Fragkoudi2017}. In \citet{Neumann2020} we report first evidence of such a separation both in cosmological simulations as well as in observations. If this happens, this would lead naturally to a more metal-rich population closer to the bar major axis.

        (ii) In the first scenario stellar populations that existed before the bar formation are separated, but it is not yet understood if this separation effect can continue after bar formation. However, there are a couple of other processes that can also produce the observed chemical disparities. After bar formation, cold gas within the inner disc is quickly \citep[$\sim 1\,$Gyr,][]{Donohoe-Keyes2019} redistributed towards the central Kpc leading to a fairly sudden quenching of the inter-bar region as seen in the star formation histories from observations \citep{James2009,James2016} and simulations \citep{Donohoe-Keyes2019}. Such a process will shut off the metal-enrichment quickly. At the same time, as long as there is gas supplied from the outer disc, gas will get funnelled to the bar ends, locations of increased star formation activity \citep{Renaud2015b,DiazGarcia2020,FraserMcKelvie2020a,Maeda2023} and stars will form there or along the bar and likely stay on bar orbits. Thus, a continued and longer chemical enrichment compared to the inter-bar region is present.

        (iii) Recent studies of spatially resolved stellar populations in nearby galaxies \citep[e.g.][]{GonzalezDelgado2014b,Zhuang2019,Zibetti2020,Sanchez2020,Neumann2021} have reported the existence of a local counterpart (r$\Sigma_\star$ZR) to the well-known global stellar mass-stellar metallicity relation (MZR), also confirmed in mock observations of the TNG50 simulation \citep{Nanni2023}. On kpc-scales, deeper potential wells in high-density regions have increased capabilities to retain outflows from stellar winds and supernovae, thus, facilitating efficient recycling and increasing the metal content of subsequent populations of stars. As seen in the results of this work, stellar bars have clearly enhanced stellar mass surface densities. Simply applying the r$\Sigma_\star$ZR leads to the observed metal difference reported here.

        (iv) A bar is a confined structure of stars on elongated quasi-periodic and periodic orbits \citep{Contopoulos1980}. The classical picture from orbital analysis in a barred galaxy potential shows a superposition of orbits in the bar of different elongation and apocentres, an enhanced orbital mixing in the bar compared to the inner disc \citep{Binney1987}. If we assume an initially axisymmetric disc with a negative metallicity gradient and that the formation of the bar introduces orbital mixing of subpopulations with different kinematics, this leads naturally to a flattening of the metallicity gradient along the bar, where high-metallicity stars from the central region may now reach larger galactocentric radii at their apocentre.
        
        
        Our finding of a larger metallicity difference in the bar region with increasing galaxy stellar mass can be linked -- on the one hand -- to the global MZR and a more efficient enrichment in more massive galaxies, and on the other hand to the downsizing of galaxy formation and bar formation: More massive galaxies have formed earlier \citep{Thomas2005,Thomas2010} and bars are likely to form earlier in more massive galaxies \citep{Sheth2012a}. Bars are likely to be more mature and strong, enhancing the processes discussed above in time, strength, and efficiency. 

        The metallicity trend with stellar mass may also be linked to a change of regime around $\log_{10} (M_{\odot}/M_{\sun}) =10$ in barred galaxies as reported from isolated hydrodynamical simulations in \citet{Verwilghen2024} and linked to observational findings in \citet{FraserMcKelvie2020a}. At higher masses, star formation is happening mostly within the central kpc in strong starbursts forming a central ring or disc, while at lower masses star formation is spread out along the bar without the formation of a central ring or disc structure. We observe a change in the metallicity distribution with a similar transition mass threshold. Higher-mass galaxies have clearly higher metallicity in the bar than in the disc, while lower-mass galaxies show no clear difference between the bar compared to the inter-bar, indicating that bars in low-mass galaxies do not significantly impact star formation and chemical enrichment processes in the interstellar medium.

    \subsection{Asymmetry of metallicity variations}

        In the sections above, we report a slight angular asymmetry between the enhanced stellar metallicity along the bar and the mass surface density, i.e. the bar itself. The asymmetry disappears if the direction spiral arms are pointing (interpreted as the sense of rotation) remains random in our averaging process. Therefore, a correlation between the angular offset and the rotation of the galaxy or the location of the spiral arms can be assumed. Enhanced metallicity is found along the bar shifted towards the trailing edge of the bar. This effect is seen particularly in star-forming bars in low-mass galaxies. There is also a tentative correlation with younger populations with similar angular distribution. We attribute this to the recent formation of young, metal-rich stars. The imprinted asymmetry in the average stellar population maps is not straight forward to explain, as orbital time periods are short ($\sim 100\,$Myr), and stars are mixed quickly in the bar. Star formation in bars usually happens on the leading edge in or close to the bar lanes, which are visible in dust absorption and molecular gas observations \citep[e.g.][]{Sheth2002a, Neumann2019, FraserMcKelvie2020a}. Therefore, the youngest star clusters are expected to be found on the leading edge. However, simple stellar population models with optical spectra are very limited when fitting young deeply embedded stellar regions. These newly formed populations are more likely to be picked up in the fit a few tens of Myr later, when they have already travelled to the trailing edge after half a radial oscillation leaving an imprint in form of increased metallicity and younger ages. 

        Another scenario assumes that star formation processes in low-mass bars are similar to those in spiral arms. Increased gas-phase metallicity has been observed on the upstream side of spirals ($=$ trailing edge when inside corotation) and explained by a simple two-phase process, in which fast mixing of gas in the spiral arm with new low-metallicity gas dilutes the gas cell when passing from the trailing to the leading edge \citep{Ho2017,Kreckel2019}. Whether this would be observable in the stellar populations is unclear at this point. A deeper investigation into this topic is outside the scope of this paper and requires data of higher spatial resolution and an individual-galaxy approach.
        


        
    \subsection{Metals beyond the bar radius}
    \label{sect:discuss_Zbeyond}

        In quenched galaxies and, in particular, galaxies with long bars (defined as $L_\mathrm{bar}/R_{90}>0.45$), we observe a metal-excess aligned with the bar major axis, but beyond the bar radius. As our polar maps show only relative differences within each radial ring, this observation is equivalent to having metal-poorer stars along the imaginary prolonged bar minor axis outside the bar radius. In the absence of spiral arms in quenched galaxies, if these features aligned with the position angle of the bar persist over time, it suggests that these stars are on orbits corotating with the bar. Hence, we speculate that what we see are metal-poor stars orbiting the L4 and L5 Lagrange points on both sides of the bar. These Lagrange points are stable points at the corotation resonance. Close to these points banana-shaped long-period orbits are elongated parallel to the bar \citep{Contopoulos1989}. These regions could have been subjected to early cessation of star formation, while it was still ongoing at the spiral arm-bar connection points. Slightly older stellar population ages in these regions yield support to this scenario. Detailed hydrodynamical simulations that run until complete gas consumption are needed to test this hypothesis. At the same time, knowledge of the galaxies' rotation curves and bar pattern speeds on an individual basis will help locate the resonance orbits.

    \subsection{Stellar ages of bars}

        Azimuthal differences in the age of stellar populations in barred galaxies are not as clear and invariable as the metallicities. Using MILES stellar population models for the fitting, we observe on average younger populations along bars compared to inter-bar regions, with the largest contrast in star-forming bars, as expected. However, across all samples, there is always a significant fraction of bars with older populations. The distribution of age contrasts is very broad with 0.1\,dex - 0.15\,dex between the first and third quartiles. In the case of using the MaStar stellar population models, which we discuss in detail in Appendix \ref{apx:ssp} and Fig. \ref{fig:violins_ssps}, we find on average older stellar populations along the bar major axis, except for star-forming bars in low-mass galaxies, again with a broad distribution. While there is a systematic shift between the results from both libraries, we can conclude that either way there is no clear answer to how the stellar ages compare between bars and inter-bar regions.
        

        Bars have been shown to be peculiar environments when it comes to star formation. Many bars show no or little sign of ongoing star formation despite continued gas inflow \citep{Neumann2019,FraserMcKelvie2020a,DiazGarcia2020,Maeda2023}. The exact conditions or evolutionary time scales that separate star-forming from non-star-forming bars are not yet clear. In the inter-bar region, on the other side, we know that star formation is often quenched relatively quickly within 1-2$\,$Gyr \citep{Donohoe-Keyes2019}. Which region ends up showing older or younger stellar populations will depend on the exact star formation history, which is difficult to assess without detailed case-by-case studies. In addition, higher spatial resolution data might help to separate highly-elongated orbits along the bar better from the disc.



    \subsection{Impact on galaxy evolution}

        A good knowledge of the distribution of stars, their age and chemical composition across cosmic time alongside with their dynamics is of utmost importance for studying galaxy evolution. Bars, which alter the inner gravitational potential substantially, are one of the key morphological structures that impact dark matter, gas and stars in a galaxy not only from a dynamical perspective but they also redistribute matter and alter star formation processes. Recent JWST and submilimeter observations have shown that galaxies form bars at very early stages, potentially already starting within the first 2$\,$Gyr \citep{Guo2023,Costantin2023,Smail2023,Amvrosiadis2024,Guo2024,LeConte2024,Tsukui2024}, highlighting the important and long-range impact of bars on the evolution of galaxies.

        In this work we present a statistical exploration of the distribution of stellar populations in a large population of nearby galaxies. We find that bars are dense and metal-rich stellar structures, whose stellar population age varies from galaxy to galaxy.

        While it is a bit of a circular outcome that bars are of higher stellar mass surface density than their surrounding discs, because it is directly connected to their definition, it is, firstly, a confirmation of the visual detection and automatic fitting method and, secondly, it offers a metric to characterise bars tightly related to their strength and gravitational pull. By analysing different subsamples, we find that the $\Sigma_\star$ contrast increases with galaxy mass (22\% on average), distance from the SFMS (19\%) and scaled bar length (13\%). This means that bars are not necessarily -- or not only -- more massive in these galaxies, but that they trap or form a larger fraction of stars within the bar radius. Thus, it is not a pure scaling effect but a change in their time-integrated impact on the host galaxy. Either long bars and bars in massive and quenched galaxies are older and have had more time to alter the stellar distribution or they are stronger or both.

        We also find that stellar populations in bars are more metal rich than in the inter-bar region. In the previous sections we discussed different scenarios that are able to produce such an outcome. They all have in common that either way bars quickly (judged by the frequency of the metal-rich feature) change how and where chemical evolution proceeds in the inner parts of galaxies: Stellar orbits are changed whereby kinematically cold and metal-rich stars might be trapped on more elongated orbits than kinematically hot and metal-poor stars, most of the inner region gets cleared of gas, which often accumulates within a central and outer rings, followed by increased star formation in these rings, chemical enrichment within the high-surface-density bar environment is likely increased, and finally, as a secondary observational effect, metallicity gradients are flattened along the bar. In summary, star formation and chemical enrichment is happening in a more localised way and measurements averaged over large areas or radial profiles are becoming less meaningful in galaxies with such environments.

        We have further seen that bars not only impact the distribution within the bar radius but also outside likely due to bar resonances similarly to gas accumulation and star formation in outer rings in barred galaxies. As for mass surface density, also the stellar metallicity contrasts show that processes in low-mass galaxies are different or much weaker. Further investigation with a high-resolution case-by-case study will shed further light into these types of bars.

        It is a poorly understood process and still remains an open question how, where and under which local conditions star formation occurs within bars \citep{Phillips1996, Verley2007c, Emsellem2015, Neumann2019, FraserMcKelvie2020a, DiazGarcia2020, Maeda2023, Kim2024}. Naturally, this is tightly connected to the build-up of metals in new generations of stars. The results of this paper that bars are metal-rich environments provide an anchor point for upcoming theoretical and simulation studies, which are required to advance towards a complete picture of the physical processes that drive the evolution of barred galaxies.

\section{Conclusions}
\label{sect:conclusion}

In this paper, we present azimuthal differences of light-weighted stellar populations in $\sim 1\,000$ barred galaxies from the MaNGA survey. The large sample size and parameter coverage allows us to stack galaxy stellar population properties and to compare between subsamples probing different galaxy and bar properties. 

As part of the analysis in this paper, we use a new method to fit galaxy bars by combining bar masks from the citizen science project Galaxy Zoo: 3D with ellipse fitting of the deprojected masks. We provide bar length measurements for 977 MaNGA galaxies. We refer to future large surveys such as Euclid, where this method can be adopted to study bars with redshift.

Our main results can be summarised as follows:

\begin{enumerate}[label={\arabic*.}]

    \item Bars are of higher stellar mass density by $0.113^{+0.065}_{-0.067}$\,dex\footnote{Quoted are the median and the interquartile range.} and are more metal-rich by $0.028^{+0.033}_{-0.040}$\,dex than the inter-bar region in the inner disc. The difference increases with total stellar mass and quenching state, likely as an effect of the presence of more mature and strong bars in more massive and gas-poorer galaxies.

    We propose a combination of different mechanisms to cause the metallicity difference including (i) kinematic separation during bar formation trapping metal-rich stellar populations on more elongated orbits, (ii) quicker quenching of star formation in the inter-bar region than in the bar, (iii) mass density-driven efficient enrichment in bars and (iv) radial mixing leading to a flatter gradient along the bar.
    
    \item The high-metallicity signature along the bars is asymmetrically offset towards the trailing edge of the bar, in particular in star-forming bars in low-mass galaxies. We discuss as possible origin either gas dilution as gas in the inner disc passes through the bar from the upstream to the downstream region similar to proposed star formation processes in spiral arms, or young, metal-rich star formation within the bar lanes with a time-delay effect being picked-up in the stellar population fit on the trailing edge.

    \item In quenched galaxies and especially in galaxies with long bars, we find a curious continued trend of metal-richer populations aligned with the bar but extending beyond the bar radius, concurrent with metal-poorer populations parallel to the bar and beyond the bar radius. We suggest a connection to metal-poor stellar populations orbiting around the L4 and L5 Lagrange points.

    \item The azimuthal age distribution of stellar populations presents a less clear trend. On average we find younger ages by $-0.026^{+0.057}_{-0.051}$\,dex employing the MILES stellar population models, especially in low-mass galaxies and star-forming bars, but with broad distributions and a large fraction of galaxies showing older ages. If we use MaStar models, the age contrasts between bar and inter-bar shifts to on average older ages in the bars ($0.021^{+0.068}_{-0.076}$\,dex), except for star-forming bars. We conclude that there is no single answer for the stellar age distribution in barred galaxies, the star formation histories need to be analysed on a case-by-case basis.

    \item In addition to the analysis of azimuthal stellar population trends, we report an interesting \emph{average} or \emph{dominant} spiral arm signature when stacking stellar mass density maps. With only aligning bars and the sense of rotation and while we not require spiral arms to connect to the bar, we observe an spiral-like overdensity connecting to the bar end and winding outwards with a pitch angle of $\alpha=25.2\degr$. This could point to a preferred bar-spiral arm connection and pitch angle (Fig. \ref{fig:polar_example}).

    \item We observe a tentative, subtle hint possibly stemming from radial migration in form of lower metallicity stars on the leading side of spiral arms and higher metallicity stars on the trailing side, as expected from theory and simulations. Alternatively, kinematic separation of different disc populations can produce similar effects.

\end{enumerate}

This work provides a statistical exploration of azimuthal variations of stellar populations in barred galaxies and constitutes a reference for both simulations, detailed studies of the Milky Way, as well as future detailed stellar population studies of low and high-redshift barred galaxies.

\section*{Acknowledgements}

We would like to thank the referee for carefully reading our paper and providing constructive comments for its improvement. JN acknowledges funding from the European Research Council (ERC) under the European Union’s Horizon 2020 research and innovation programme (grant agreement No. 694343).

Numerical computations were done on the Sciama High Performance Compute (HPC) cluster which is supported by the ICG, SEPnet and the University of Portsmouth.

Funding for the Sloan Digital Sky Survey IV has been provided by the Alfred P. Sloan Foundation, the U.S. Department of Energy Office of Science, and the Participating Institutions. SDSS acknowledges support and resources from the Center for High-Performance Computing at the University of Utah. The SDSS web site is \url{www.sdss.org}.

SDSS is managed by the Astrophysical Research Consortium for the Participating Institutions of the SDSS Collaboration including the Brazilian Participation Group, the Carnegie Institution for Science, Carnegie Mellon University, the Chilean Participation Group, the French Participation Group, Harvard-Smithsonian Centre for Astrophysics, Instituto de Astrof\'isica de Canarias, The Johns Hopkins University, Kavli Institute for the Physics and Mathematics of the Universe (IPMU) / University of Tokyo, the Korean Participation Group, Lawrence Berkeley National Laboratory, Leibniz Institut f\"ur Astrophysik Potsdam (AIP), Max-Planck-Institut f\"ur Astronomie (MPIA Heidelberg), Max-Planck-Institut f\"ur Astrophysik (MPA Garching), Max-Planck-Institut f\"ur Extraterrestrische Physik (MPE), National Astronomical Observatories of China, New Mexico State University, New York University, University of Notre Dame, Observatório Nacional / MCTI, The Ohio State University, Pennsylvania State University, Shanghai Astronomical Observatory, United Kingdom Participation Group, Universidad Nacional Aut\'onoma de M\'exico, University of Arizona, University of Colorado Boulder, University of Oxford, University of Portsmouth, University of Utah, University of Virginia, University of Washington, University of Wisconsin, Vanderbilt University, and Yale University.

\section*{Data Availability}

Bar length measurements are available as supplemental online material of the journal and it will be made available on ViziR. The \texttt{MaNGA {\sc firefly} VAC} described in this paper can be downloaded from the SDSS website \url{https://www.sdss.org/dr17/manga/manga-data/manga-firefly-value-added-catalog} or from the ICG Portsmouth's website \url{http://www.icg.port.ac.uk/manga-firefly-vac}. It is also available as {\sc cas} table on the SDSS skyserver \url{http://skyserver.sdss.org/dr17} and integrated in {\sc marvin} \url{https://dr17.sdss.org/marvin}. The {\sc firefly} code is publicly available at \url{https://github.com/FireflySpectra/firefly_release} and is described at \url{https://www.icg.port.ac.uk/firefly} and \url{https://www.sdss.org/dr17/spectro/galaxy_firefly}. The stellar population models used in this paper are available at \url{https://svn.sdss.org/public/data/sdss/stellarpopmodels/tags/v1_0_2/} and \url{http://www.icg.port.ac.uk/mastar} and also integrated in the {\sc firefly} github package.



\bibliographystyle{mnras}
\bibliography{NewDatabase} 




\appendix

\section{Differences between using MILES or MaStar SSPs}
\label{apx:ssp}

    \begin{figure*}
        \centering
        \includegraphics[width=\linewidth]{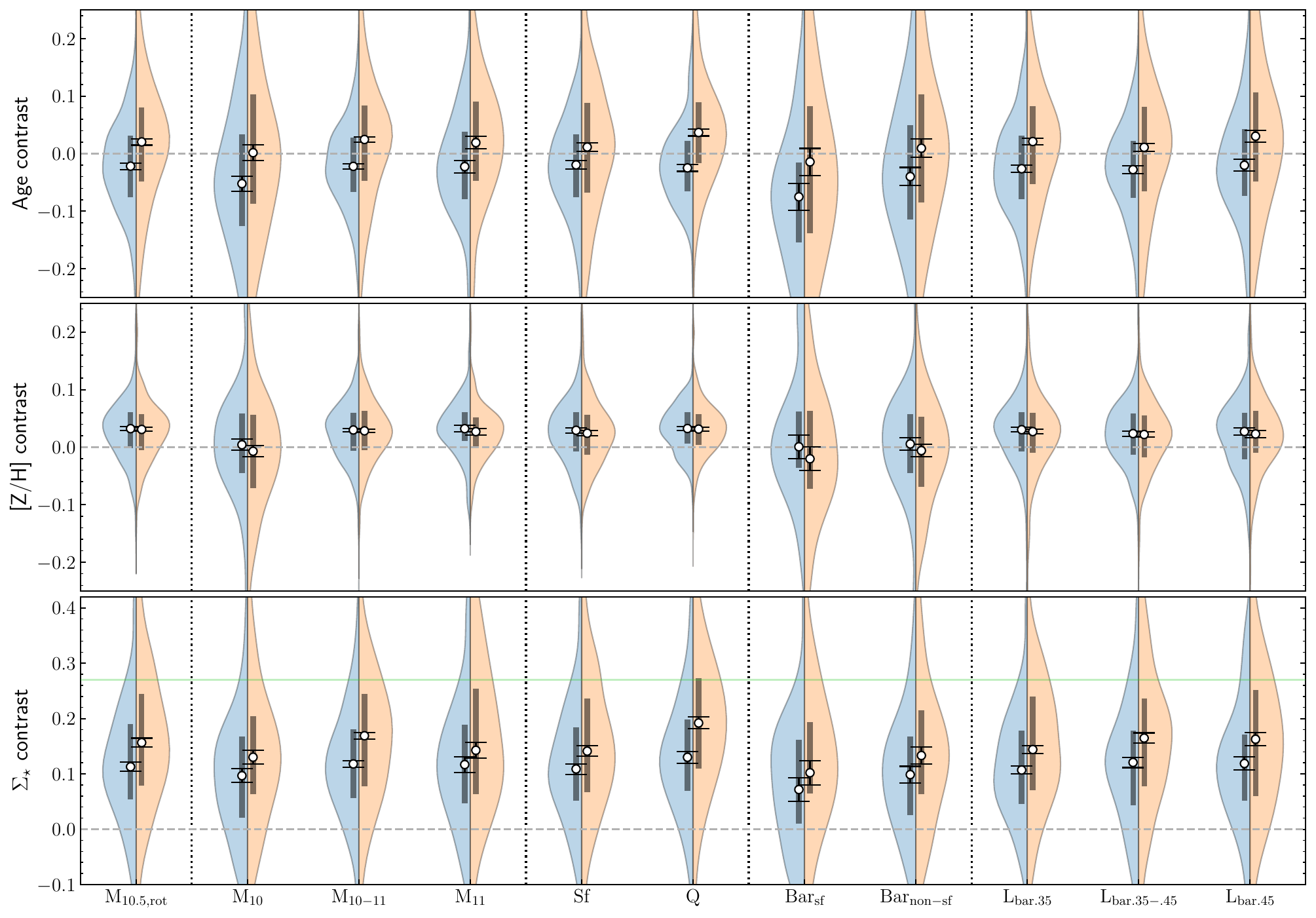}
        \caption{Similar to Fig. \ref{fig:violins} but showing the different distributions when either using MILES SSPs (left) or MaStar SSPs (right). The green solid line shows the $\Sigma_\star$ contrast of the Milky Way bar using the analytical fit presented in \citet{Sormani2022} of the dynamical Milky Way model by \citet{Portail2017a}.}
        \label{fig:violins_ssps}
    \end{figure*}

    \begin{figure}
        \centering
        \includegraphics[width=\columnwidth]{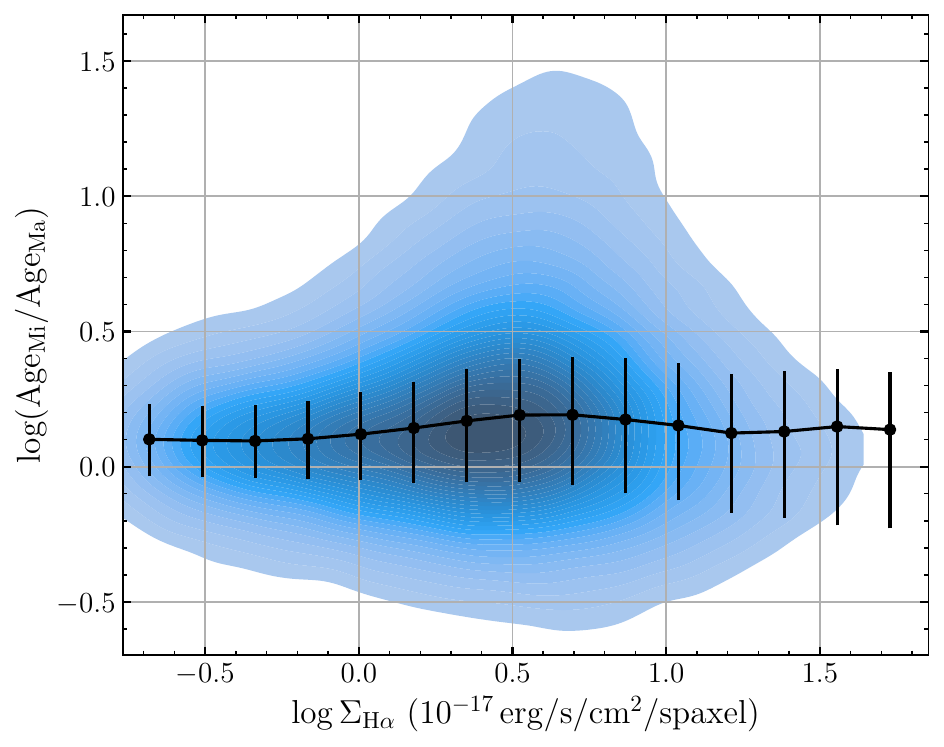}
        \caption{Differences in mean stellar ages between the fits using the MILES and MaStar library and their dependence on H$\alpha$ surface density as obtained from \citet{Westfall2019} and \citet{Belfiore2019}.}
        \label{fig:age_diff}
    \end{figure}

We test the effect of changing the stellar population model library when fitting our data. In Fig. \ref{fig:violins_ssps} we show how the main results presented in Fig. \ref{fig:violins} compare between employing both variants of the \ffvac{}. Noticeably, there is no difference between all of the metallicity contrasts within the error bars, corroborating our main result. The stellar age contrast, on the other hand, is shifted systematically to higher values when using MaStar compared to lower values when using MILES, such that the median age contrast is positive in MaStar meaning older populations along the bar than the inter-bar region. Yet, the distributions of these numbers between both datasets show a significant overlap and in both cases there are always at least 30\% of galaxies either with older or younger populations. The conclusion remains that there is no single answer to the question whether bars have older or younger populations than inter-bar regions.

It is not easy to assess which of the libraries provides the closer solution to the true average age trend. In \citet{Neumann2022}, we conducted a series of tests to understand the differences. The largest differences between fitted stellar population parameters between both \ffvac{} variants are indeed in the stellar ages. The most convincing explanation is the lack of young templates in the MILES library, as a result MaStar fitted ages are on average younger. In Fig. \ref{fig:age_diff}, we test how the age difference depend on SFR traced by H$\alpha$ emission. As expected, the higher the H$\alpha$ surface density, indicating larger fractions of very young stars, the larger the difference in ages from the stellar population fit.

The age contrast in Fig. \ref{fig:violins_ssps} only compares relative azimuthal variations. We looked into the absolute age values along the bar and inter-bar and found that in both cases MaStar finds younger ages compared to MILES, but the discrepancy is larger in the inter-bar region. Thus the bar to inter-bar age contrast becomes older. This is directly affecting the fitted mass surface density. Older populations are less luminous, therefore, more mass of them is needed to make up for the flux deficit. Hence, we observe higher $\Sigma_\star$ contrasts in MaStar compared to MILES. Using the stellar mass surface density variation between bar and inter-bar might be a way to understand which library provides better results for the stellar ages.

A general indication can be provided by dynamical models of galaxies, such as the observationally well-constrained Milky Way model in \citet{Portail2017a}. We use the analytical fit presented in \citet{Sormani2022} to obtain a surface density distribution across the face-on view of the Milky Way. Furthermore, we fit ellipses to iso-density contours to obtain the length of the bar following a similar procedure as in Sect. \ref{sect:lbar} and find $L_{\rm bar,\ MW}=5.68\,$kpc in relatively good agreement with $5.0 \pm 0.2$\,kpc in \citet{Wegg2015} and $5.30 \pm 0.36$\,kpc in \citet{Portail2017a}. Finally, we compute the mass surface density contrast in equally defined bar and inter-bar regions and obtain a $\Sigma_\star$ contrast of 0.27\,dex, shown in Fig. \ref{fig:violins_ssps}. Compared to our MaNGA results this number is relatively large and within the upper 25\% of both distributions. At the same time, it gives a general indication that the MaStar library might work better.


\section{Impact of galaxy inclination}
\label{apx:test_incl}

    \begin{figure}
        \centering
        \includegraphics[width=\columnwidth]{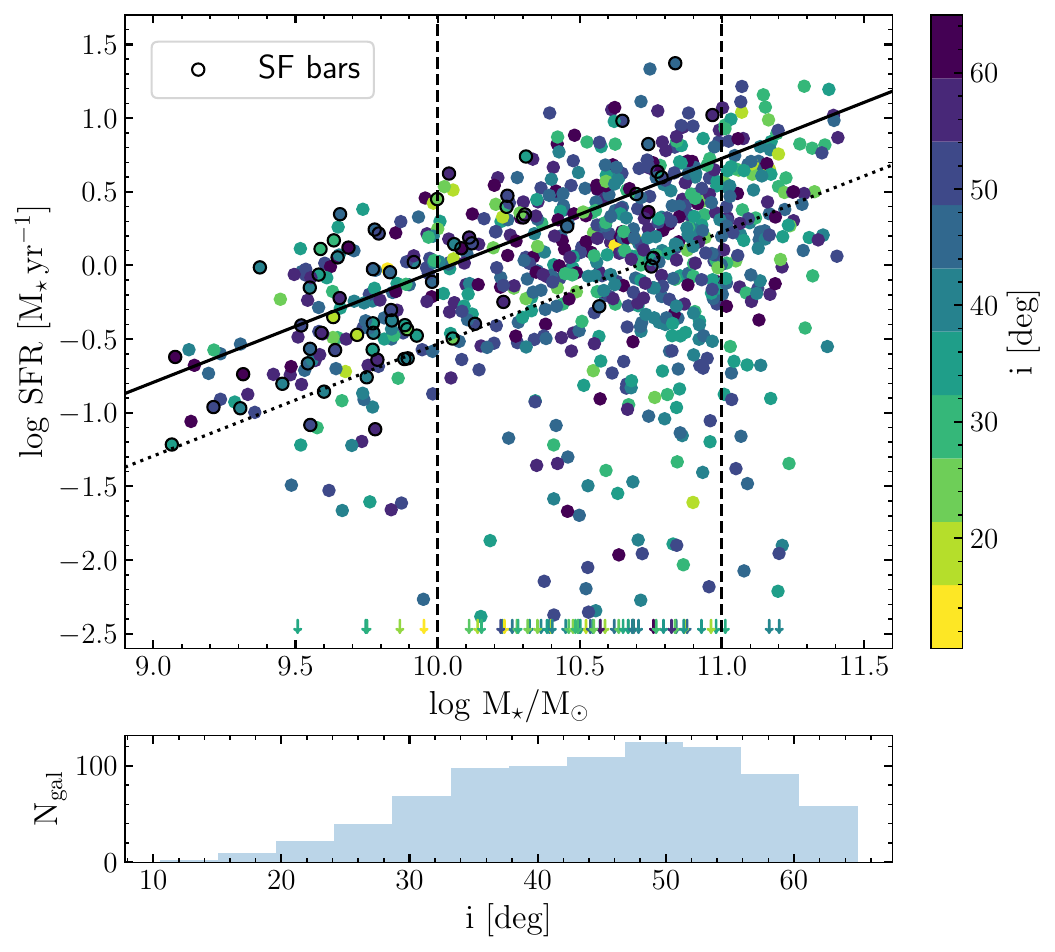}
        \caption{\textit{Top panel:} same as Fig. \ref{fig:sample} but coloured by galaxy inclination. \textit{Bottom panel:} shows a histogram of the inclinations.}
        \label{fig:incl_test_sample}
    \end{figure}

    \begin{figure}
        \centering
        \includegraphics[width=\columnwidth]{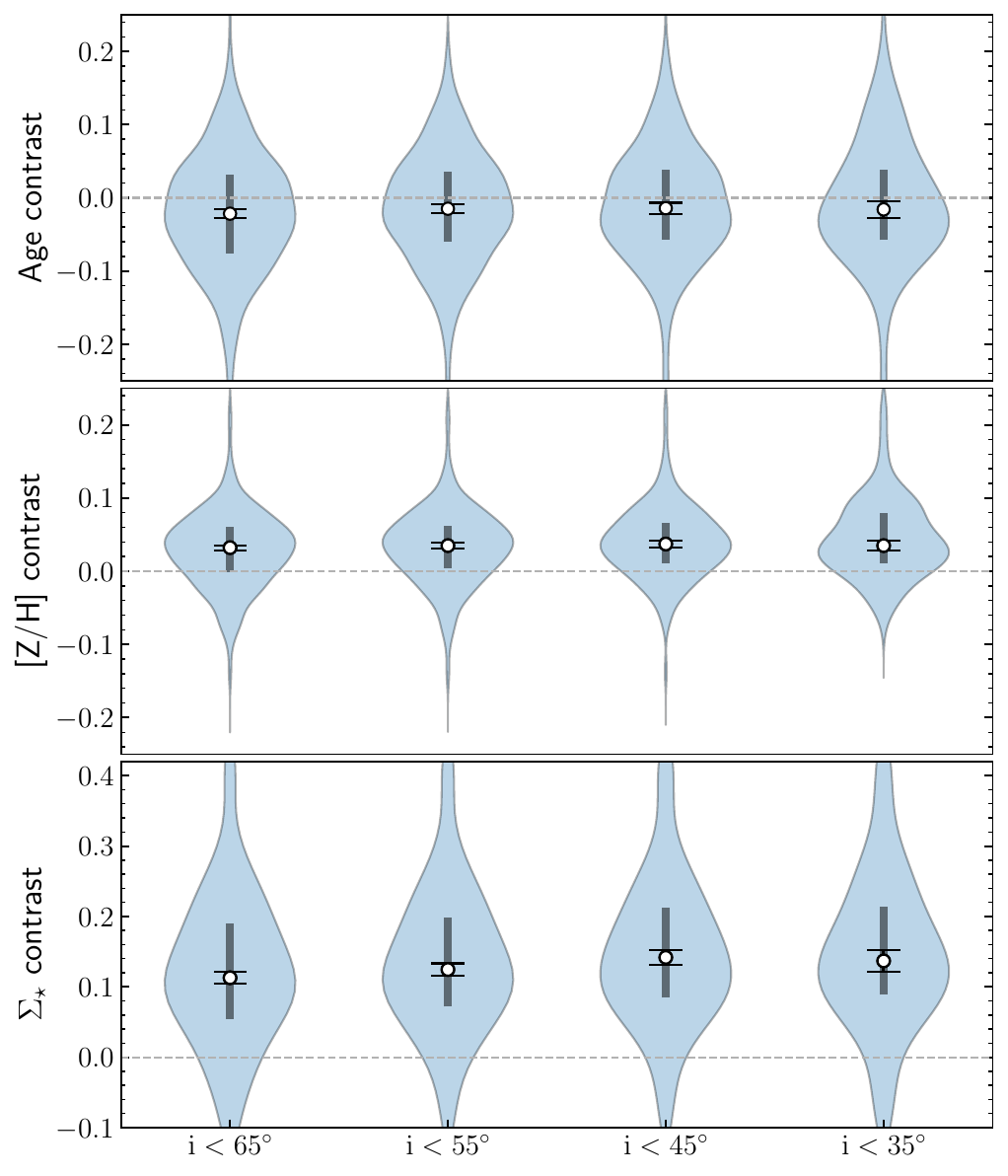}
        \caption{Same as Fig. \ref{fig:violins} but here showing the $\rm M_{10.5, rot}$ sample for four different inclination cutoffs, from left to right with 374, 305 ,184 and 92 galaxies in each sample.}
        \label{fig:incl_test_sample2}
    \end{figure}

    We test the robustness of our results against the effect of galaxy inclination. In our analysis all galaxies are geometrically deprojected under the assumption that the disc and the bar are essentially flat. Our sample excludes a priori highly inclined galaxies with $i>65\degr$. However, galaxies with intermediate inclinations might introduce uncertainties in our results due to the 3D nature of the bar.
    
    In Fig \ref{fig:incl_test_sample}, we repeat a version of our sample figure, Fig. \ref{fig:sample}, but with galaxies coloured by their inclination. In the bottom panel, we add a histogram of all galaxy inclinations. No clear biases are introduced by a potential inclination effect in our selection of subsamples, but sample size is reduced quickly if we were to enforce stricter cutoff limits.
    
    We repeat our stellar population analysis with the inclination limited to be smaller than 65$\degr$, 55$\degr$, 45$\degr$ and 35$\degr$. The main results reported in this work do not change. As an example, in Fig. \ref{fig:incl_test_sample2}, we show the distribution of the $\Sigma_\star$, $[Z/H]$ and age contrasts for our subsample of galaxies with known rotation and $\log (M_\star/M_\odot)>10.5$ from Sect. \ref{sect:results_general}. Changes of the medians of age and [Z/H] are within the errorbars with a small shift of the distribution towards a higher metallicity contrast for the most face-on galaxies. The mass surface density contrast increases also slightly with more stringent inclination limits. At the same time, the sample size is reduced from 374 to 92 galaxies. We conclude that we can safely trust the results of this work in the context of deprojection effects.

\section{Additional data}
\label{apx:data}

    \begin{table}
	\caption{Bar length measurements. The first 10 entries are shown here, the full table of 976 galaxies is available online.}
	\label{tbl:bar_lengths}
	\centering
        \resizebox{\columnwidth}{!}{%
	\begin{tabular}{rrrrrrrr} 
		\hline
Internal ID & PLATE-IFU &   $L_{\rm bar}$ &  $Ell_{\rm bar}$ & $PA_{\rm bar}$\\
 & & SMA (arcsec) & & (deg)\\
		\hline
0	& 10001-1902	& 5.88	& 0.68	& 162.87\\
1	& 10001-6102	& 8.86	& 0.56	& 17.99\\
2	& 10001-6103	& 6.78	& 0.60	& 179.16\\
3	& 10001-9101	& 4.24	& 0.51	& 156.89\\
4	& 10213-12705	& 9.08	& 0.31	& 7.84\\
5	& 10215-3704	& 3.71	& 0.62	& 167.94\\
6	& 10216-6101	& 4.62	& 0.65	& 19.69\\
7	& 10217-12705	& 6.29	& 0.63	& 40.40\\
8	& 10217-6103	& 5.48	& 0.48	& 7.20\\
9	& 10218-12702	& 3.13	& 0.54	& 171.87\\
            \hline
	\end{tabular}}
    \end{table}	    


\bsp	
\label{lastpage}
\end{document}